\documentclass[a4paper,twocolumn,aps,prx]{revtex4-2}

\newcommand{\eq}[1]{\begin{equation}#1\end{equation}}
\newcommand{\dd}{\mathrm{d}}
\newcommand{\ee}{\mathrm{e}}

\newcommand{\tea}{\tilde \varepsilon_a}
\newcommand{\teb}{\tilde \varepsilon_b}

\newcommand{\Tr}{\mathrm{Tr\,}}
\newcommand{\Trb}{\mathrm{Tr}_B}
\newcommand{\rp}{\mathrm{Re}}
\newcommand{\ip}{\mathrm{Im}}

\newcommand{\lneg}{\mathcal{E}}

\newcommand{\Li}[1]{\mathrm{Li}_2\left( #1 \right)}

\newcommand{\twomat}[4]{\left(\begin{array}{cc} #1 & #2 \\ #3 & #4\end{array}\right)}
\newcommand{\identity}{\openone}

\newcommand{\const}{\mathrm{const}}

\usepackage{amsmath}
\usepackage{amsfonts}
\usepackage{graphics}
\usepackage{graphicx}
\usepackage{psfrag}
\usepackage{color}
\usepackage{colordvi}
\usepackage{bbm}

\usepackage{hyperref}

\begin{document}

\title{Entanglement negativity in a nonequilibrium steady state}

\author{Viktor Eisler}

\affiliation{
Institute of Theoretical and Computational Physics,
Graz University of Technology, Petersgasse 16, 8010 Graz, Austria}

\begin{abstract}
We study entanglement properties in a nonequilibrium steady state of a free-fermion chain,
that emerges after connecting two half-chains prepared at different temperatures.
The entanglement negativity and the R\'enyi mutual information between two adjacent
intervals scale logarithmically in the system size, with prefactors that we calculate analytically
as a function of the bath temperatures. In particular, we show that the negativity and the
R\'enyi mutual information with index $\alpha=1/2$ are described by different prefactors,
and thus the two quantities provide inequivalent information about the state.
Furthermore, we show that the logarithmic growth of the negativity during time
evolution is also governed by the steady-state prefactor.

\end{abstract}

\maketitle

\section{Introduction}

The study of entanglement properties in many-body systems had a vast contribution
to our understanding of quantum phases of matter \cite{AFOV08,CCD09,Laflo16}.
In particular, the structure of entanglement in pure ground states of lattice systems
with local interactions is characterized by an area law \cite{ECP10}. In one-dimensional
systems, logarithmic violations of the area law signal the criticality of the state,
with a prefactor that is universal and described by conformal field theory (CFT) \cite{CC09}.

Despite the enormous progress for pure states, the characterization of entanglement in
mixed states remains a challenge. The core of the problem is finding an efficiently computable
measure of entanglement, with a prospective candidate being the logarithmic negativity
\cite{EP99,VW02,Plenio05}. In the quantum field theory framework, the entanglement negativity between
two segments can be evaluated in ground \cite{CCT12,CCT13,BFCAD16} and thermal states
\cite{EZ14b,CCT14,SR19b} of a 1D chain. These predictions are nicely confirmed by numerical calculations
in harmonic and spin chains \cite{AEPW02,AW08,FCGA08,MRPR09,WMVB09,WVB10,CTC13,SDHS16}.
Results for higher dimensional lattices also exist \cite{LV13,Castelnovo13,EZ16,DNCT16},
and the role of negativity in finite-temperature phase transitions has been addressed \cite{WAA20,WLCKG20}.

The studies have also been extended to the out-of-equilibrium scenario, and the
time evolution of entanglement negativity was investigated in quantum quenches \cite{CTC14,WCR15,FG19,PBC22}.
Interestingly, for free-particle chains it has been observed that there is an intimate proportionality
relation between the negativity and the R\'enyi mutual information (RMI) with index $\alpha=1/2$ \cite{AC19}.
Namely, to leading order one finds that the two quantities are related by a factor two, which
has been substantiated in a number of other quench scenarios \cite{GE20,BKL22,RT22,FG22}.
In particular, the results of \cite{BKL22} suggest that the relation should hold for rather generic
unitary dynamics, provided that the initial state is pure.

On the other hand, recent counterexamples were found in the context of non-unitary dissipative
dynamics, where the leading order behaviour of negativity and RMI were found to be different \cite{AC22,CA22}.
However, this cannot be a completely generic property of mixing dynamics, as the proportionality
has been recovered in other non-unitary cases \cite{TPS22} and even for random mixed states \cite{SLKFV21}.
Furthermore, the inequivalence of entanglement negativity and RMI has also been pointed out
in quench dynamics of irrational CFTs \cite{KFKR20,KFKR21}. The above examples clearly demonstrate,
that the relation between these two quantities is far from being understood.

In this paper we shall provide an example for a mixed state of a free-fermion chain,
where the relation between negativity and RMI breaks down. Namely, we consider
a nonequilibrium steady state (NESS) that emerges under unitary dynamics, after two
half-chains at different temperatures are joined together. In the very same setting,
the standard mutual information was shown to violate the area law \cite{EZ14},
and similar logarithmic violations have been found in the NESS of various other
free-fermion related chains \cite{ABZ14,Ribeiro17,KZ17,FG21,AAR22}. In sharp contrast,
in the analogous NESS of a harmonic chain both the negativity and the mutual information
were found to satisfy the area law \cite{EZ14b}, which is the result dictated
by free-boson CFT \cite{HD15}.

Here we study the RMI with $\alpha=1/2$ as well as the negativity between two adjacent
or disjoint intervals in the fermionic NESS, and show that they both scale logarithmically in the
segment size or cross ratio, respectively. However, the prefactors are described by different
functions of the temperatures, and hence the two quantities provide inequivalent information
about the NESS. Our results are obtained from the R\'enyi generalization of the negativity
via an analytic continuation, using the techniques introduced in \cite{FG22}.
The analytical results are in perfect agreement with our numerical calculations.
Furthermore, we also provide numerical evidence that the negativity between adjacent
intervals grows logarithmically in time during the dynamics, governed by the exact same
prefactor found for the NESS.

The rest of the manuscript is structured as follows. In Section \ref{sec:model} we
introduce the model and describe the setup. The results for the R\'enyi mutual
information as well as for the entanglement negativity are presented in
Sec.~\ref{sec:rmi} and \ref{sec:neg}, respectively. Our findings are discussed
in Sec.~\ref{sec:disc}, followed by two appendices with some technical details
of the calculations.

\newpage

\section{Model and setup\label{sec:model}}

We consider free fermions on an infinite chain, described by the Hamiltonian
\eq{
H=-\frac{1}{2}\sum_{m=-\infty}^{\infty}
\left(c^{\dag}_{m}c^{\phantom\dag}_{m+1}  + c^\dag_{m+1}c^{\phantom\dag}_m\right) ,
\label{H}}
with fermionic creation and annihilation operators satisfying the canonical
anticommutation relations $\{ c^\dag_m,c_n\}=\delta_{m,n}$.
We are interested in the nonequilibrium dynamics of the hopping chain
that emerges due to an inhomogeneous initial state, characterized by
two different temperatures $\beta_{l}$ and $\beta_{r}$ on the left and right side
\eq{
\rho_0 = \frac{1}{Z_{l}} \ee^{-\beta_l H_l} \otimes
\frac{1}{Z_{r}} \ee^{-\beta_r H_r} \, .
\label{rho0}}
The half-chain Hamiltonians $H_l$ and $H_r$ have the same form as
\eqref{H}, but with sums running over sites $-\infty$ to $-1$ and
$1$ to $\infty$, respectively. The state $\rho_0$ is thus a
tensor product of two Gibbs states at different temperatures,
which for $t>0$ evolves under the unitary dynamics
\eq{
\rho_t = \ee^{-i H t} \rho_0 \, \ee^{i H t} \, .
\label{rhot}}

The main focus of our studies is the nonequilibrium steady state
$\rho_\infty$, and its entanglement properties. For the initial state
\eqref{rho0} it has been shown \cite{AH00,Ogata02,AP03} that the NESS is locally well defined by
the requirement
\eq{
\Tr (\rho_\infty \mathcal{O}) = \lim_{t\to\infty} \Tr (\rho_t \mathcal{O})
}
for any observable $\mathcal{O}$ supported on a finite set of sites.
Furthermore, since $\rho_\infty$ is a Gaussian state, it is fully characterized
by its correlation matrix
\eq{
C_{mn} =
\Tr (\rho_\infty c_m^{\dag} c_n^{\phantom\dag}) =
\int_{-\pi}^{\pi} \frac{\dd q}{2\pi} \ee^{-iq(m-n)} n_q ,
\label{cmn}}
where the occupation function is given by \cite{Ogata02}
\eq{
n_q =
\begin{cases}
\frac{1}{\ee^{-\beta_r \cos q}+1} & -\pi < q < 0 \\
\frac{1}{\ee^{-\beta_l \cos q}+1} & \phantom{-}0 < q < \pi
\end{cases}.
\label{nq}}
In other words, the right-moving fermionic modes are thermalized
at the temperature of the left-hand side bath and vice-versa. 
In fact, the NESS can be understood in a simple hydrodynamic picture,
where the non-interacting modes propagate ballistically and carry the
information of their initial occupation to large distances.

The conventional way of characterizing entanglement is to consider
a bipartition of the system into a subsystem $A$ and its environment $B$,
and consider the reduced density matrix $\rho_A = \Trb (\rho_\infty)$.
However, since the NESS is a mixed state, the von Neumann entropy
\eq{
S(\rho_A) = -\Tr (\rho_A \ln \rho_A)
}
is not a proper measure of entanglement. Instead, one could consider
a tripartite scenario with subsystems $A_1$, $A_2$ and $B$, and calculate
the mutual information
\eq{
\mathcal{I}(A_1:A_2)=
S(\rho_{A_1})+S(\rho_{A_2})-S(\rho_{A_1\cup A_2}) \, ,
\label{mi}}
which is a measure of the total correlations between $A_1$ and $A_2$. For two adjacent intervals,
$\mathcal{I}(A_1:A_2)$ satisfies an area law in thermal equilibrium for arbitrary
local Hamiltonians \cite{WVHC08}.
However, for the NESS at hand it has been shown in \cite{EZ14}, that the mutual information
violates the area law and scales logarithmically in the subsystem size.
This behaviour is due to the jump singularity in the occupation
function \eqref{nq} at $q=0$, between
\eq{
a=\frac{1}{\ee^{-\beta_r}+1},\qquad
b=\frac{1}{\ee^{-\beta_l}+1},
\label{ab}}
while a second jump between the values $1-a$ and $1-b$ occurs at $q=\pm \pi$.

The mutual information quantifies only the total (quantum + classical) correlations between
two subsystems in the NESS. In the following we shall extend the calculations to obtain the
logarithmic negativity, which is a proper measure of entanglement \cite{VW02}.
As a first step towards this goal, we study the R\'enyi mutual information, which was 
shown to have a very close relation to the negativity in the context of quantum quenches \cite{AC19,BKL22}.
The techniques introduced in the next section will be directly applicable to the calculation of the
negativity.

\section{R\'enyi mutual information\label{sec:rmi}}

The R\'enyi mutual information is defined as
\eq{
\mathcal{I}_\alpha(A_1:A_2)=
S_\alpha(\rho_{A_1})+S_\alpha(\rho_{A_2})-S_\alpha(\rho_{A_1\cup A_2})\, ,
\label{rmi}}
via the R\'enyi entropy
\eq{
S_\alpha(\rho_A) = \frac{1}{1-\alpha} \Tr (\rho_A^\alpha) \, .
}
It is completely analogous to the standard mutual information \eqref{mi},
which is reobtained in the limit $\alpha \to 1$. For generic $\alpha$, however, the RMI
is not even a proper measure of correlations. Indeed, for the NESS of a transverse Ising
chain emerging from the initial state \eqref{rho0} it has been shown, that the RMI may become
negative for indices $\alpha > 2$ \cite{KZ17}. On the other hand, for $0<\alpha<2$ the subadditivity
of the R\'enyi entropy was proven for arbitrary fermionic Gaussian states, ensuring the
positivity of the RMI \cite{CLE19}.

Our main goal is to derive the asymptotics of the RMI with index $\alpha=1/2$
between two (adjacent or disjoint) intervals of size $\ell \gg 1$.
The strategy is to first express the RMI with integer index $n>1$ and then perform an
analytic continuation. Indeed, for integer values of the R\'enyi index, the entropy
can be calculated as \cite{CFH05}
\eq{
S_n(\rho_{A}) =
\frac{1}{1-n}\sum_{k=-\frac{n-1}{2}}^{\frac{n-1}{2}} \ln Z_{A,k} \, ,
\label{sn}}
where $Z_{A,k}$ is the generating function of the full counting statistics (FCS)
of the particle number $\hat N_{A}$ in $A$
\eq{
Z_{A,k} = \Tr(\rho_\infty \ee^{i\lambda_k \hat N_{A}})=
\det \left[\identity_A + (\ee^{i\lambda_k}-1)C_A \right] .
\label{Zk}}
Here $\identity_A$ is the identity matrix on subsystem $A$, and the phases must be
evaluated at the discrete values 
\eq{
\lambda_k = \frac{2\pi}{n}k \, , \qquad
k = -\frac{n-1}{2}, \dots, \frac{n-1}{2} \, .
\label{lamk}}
Note that the index $k$ takes integer/half-integer values for $n$ odd/even,
and $C_A$ is the reduced correlation matrix with elements \eqref{cmn}
restricted to $m,n \in A$. While originally obtained via
the replica trick in a field theoretic setting \cite{CFH05}, Eq. \eqref{sn} can also
be derived immediately using the determinant formula for the R\'enyi entropy of
Gaussian states, see Appendix \ref{app:det}.

In order to obtain the RMI in \eqref{rmi}, we need to evaluate the entropies
for the intervals $A_1$ and $A_2$, which involves the calculation of determinants
\eqref{Zk} of a Toeplitz matrix with a symbol \eqref{nq} that has a jump singularity \cite{BT91}.
This can be performed using standard Fisher-Hartwig techniques \cite{JK04,KM05}
as in \cite{EZ14}. However, for disjoint intervals this is not any more true for the
subsystem $A_1 \cup A_2$. To overcome this problem, one can apply a trick that
allows us to deal with Toeplitz determinants, at the expense of modifying the
occupation function \eqref{nq}. Indeed, the main argument is that the logarithmic
contribution of the RMI we are interested in should depend only on the parameters
$n_{q\to 0^-}=a$ and $n_{q\to 0^+}=b$ defined in \eqref{ab}. Therefore, we introduce
the piecewise constant occupation function
\eq{
n'_q = 
\begin{cases}
a & -\pi < q \le 0 \\
b & \phantom{-}0 < q \le \pi
\end{cases}.
\label{nqpc}}
and the corresponding correlation matrix
\eq{
C'_{ij} = \sum_{q} n'_q \, \varphi^*_q(i) \varphi^{\phantom{*}}_q(j) \, .
\label{Cpmn}}
Note that we also regularize the problem by considering sites on a ring
of finite size $L$, such that the momenta $q$ are integer multiples of $2\pi/L$
and the corresponding eigenstates are $\varphi_q(j) \sim \ee^{iqj}$ up to normalization.
Clearly, the correlation matrix \eqref{Cpmn} will not reproduce the correct extensive
part of the entropy in the NESS, which, however, anyway cancels out in the RMI.

We thus proceed with the calculation of the determinant in \eqref{Zk} by exchanging
$C_A \to C'_A$. Let us first rewrite
\eq{
C'_A = a \, C^0_A + b \, (\identity_A -C^0_A)\, ,
\label{CpA}}
where
\eq{
C^0_{ij} = \sum_{q \le 0} \varphi^*_q(i) \varphi^{\phantom{*}}_q(j) \, .
\label{C0mn}}
Then $C^0_A$ is nothing else but the reduced correlation matrix of a Fermi-sea
ground state with all the negative momenta occupied. This is known
to have the same eigenvalues as the overlap matrix \cite{Klich06}
\eq{
M^0_{pq} =  \sum_{j \in A} \varphi^*_p(j) \varphi^{\phantom{*}}_q(j) \, .
}
where the sum runs over the subsystem $A$ and the momentum
indices are restricted to the Fermi sea, $p,q \le 0$.
The FCS can thus be rewritten as
\eq{
Z_{A,k} = \det\left[\identity + M \right]
\det \left[(1-b+b \, \ee^{i\lambda_k})\identity_A\right],
\label{Zk2}}
where
\eq{
M_{pq}=\frac{(a-b)(\ee^{i\lambda_k}-1)}{1+b (\ee^{i\lambda_k}-1)} M^0_{pq} \, ,
\label{M}}
and $\identity$ is the identity over the Fermi sea in momentum space.

It is easy to see that the matrix $M$ with elements defined in \eqref{M} has a Toeplitz
structure for \emph{arbitrary} subsystems $A$. For simplicity, we consider two disjoint intervals
of equal length $A_1=\left[1,\ell \right]$ and $A_2=\left[d+\ell+1,d+2\ell \right]$, by keeping
the ratios $d/L$ and $\ell/L$ fixed. Introducing $\theta=2\pi j/L$
and taking the thermodynamic limit $L \to \infty$, one obtains the integral
\eq{
(\identity + M)_{pq} = \int_{0}^{2\pi} \phi_A(\theta) \ee^{-i(p-q)\theta} \dd \theta \, ,
\label{Mtoep}}
with a piecewise constant symbol $\phi_A(\theta)$ that depends on the
parameters $a,b$ and the phase $\lambda_k$. In particular, for a single
interval one has
\eq{
\phi_{A_1}(\theta) = \begin{cases}
\Phi(\lambda_k) & 0<\theta< \theta_1 \\
1 & \theta_1 < \theta < 2\pi
\end{cases},
\label{phi}}
where
\eq{
\Phi(\lambda)= \frac{1-a+\ee^{i\lambda} a}{1-b+\ee^{i\lambda} b} \, ,
\label{Phi}}
and the two jump singularities are located at
\eq{
\theta_0 = 0 \, , \qquad
\theta_1=\frac{2\pi\ell}{L} \, .
\label{theta1}}
On the other hand, for the composite subsystem $A_1 \cup A_2$ we have the symbol
\eq{
\phi_{A_1 \cup A_2}(\theta) = \begin{cases}
\Phi(\lambda_k) & \theta \in \left[0,\theta_1 \right] \cup \left[\theta_2,\theta_3\right]\\
1 & \mathrm{otherwise}
\end{cases},
\label{phi12}}
with additional jump locations at
\eq{
\theta_2=\frac{2\pi(d+\ell)}{L} \, , \qquad
\theta_3=\frac{2\pi(d+2\ell)}{L} \, .
\label{theta23}}

One can now apply the Fisher-Hartwig theorem to evaluate the Toeplitz determinant
in \eqref{Zk2}. Indeed, let us rewrite the symbol in the form
$\phi_A(\theta)=\prod_s g_s(\theta)$ where
\eq{
g_s(\theta) = \begin{cases}
\ee^{i\pi \beta_s} & 0 \le \theta < \theta_s \\ 
\ee^{-i\pi \beta_s} & \theta_s \le \theta < 2\pi
\end{cases},
}
and $s$ labels the FH singularities. The logarithm of the $L/2 \times L/2$
Toeplitz determinant then reads \cite{DIK11}
\begin{align}
&\ln \det (\identity+M) =  \frac{L}{2} \sum_{s} i \, \theta_s \beta_s 
+ \sum_s \mathcal{C}_s
\nonumber \\
&+ 2\sum_{s_1<s_2} \beta_{s_1}\beta_{s_2}
\ln \left[L \sin \Big(\frac{\theta_{s_2}-\theta_{s_1}}{2}\Big)\right] ,
\label{FH}
\end{align}
where the constant $\mathcal{C}_s$ depends on the jump only via $\beta_s$,
and we assumed $\sum_s \beta_s=0$.

Let us first consider the case of adjacent intervals, where both symbols
have two FH singularities with $s=0,1$. In particular, for $\phi_{A_1}(\theta)$ the jumps at
$\theta_0=0$ and $\theta_1$ are characterized by $\beta_1=-\beta_0=\beta(\lambda_k)$,
where
\eq{
\beta(\lambda_k) = \frac{1}{2\pi i}
\ln \left[\Phi(\lambda_k) \right]. 
\label{betalam}}
Applying the FH formula \eqref{FH} and adding the contribution of the
second determinant in \eqref{Zk2} then gives
\begin{align}
\ln Z_{A_1,k} &= \frac{\ell}{2} 
\ln \left[(1-a+\ee^{i\lambda_k} a)(1-b+\ee^{i\lambda_k}b)\right]
\nonumber \\
&- 2\beta^2(\lambda_k) \ln \left [L \sin ( \pi \ell/L)\right] 
+ 2 \, \mathcal{C}(\lambda_k),
\end{align}
where $\mathcal{C}(\lambda_k)$ is a subleading constant.
The result for $A_1 \cup A_2$ simply follows by substituting $\ell \to 2\ell$.
It is easy to see that the extensive term cancels out in the RMI,
and after carrying out the sum over $k$ as in \eqref{sn} and taking the
limit $L \to \infty$ one arrives at
\eq{
\mathcal{I}_n = \sigma_n \ln \ell + \const ,
}
where the prefactor reads
\eq{
\sigma_n = 
\frac{1}{1-n}\frac{1}{2\pi^2}\sum_{k=-\frac{n-1}{2}}^{\frac{n-1}{2}}
\ln^2 \left(\frac{1-a+\ee^{i\lambda_k} a}{1-b+\ee^{i\lambda_k} b}\right).
\label{sign}}

\begin{figure}[htb]
\center
\includegraphics[width=\columnwidth]{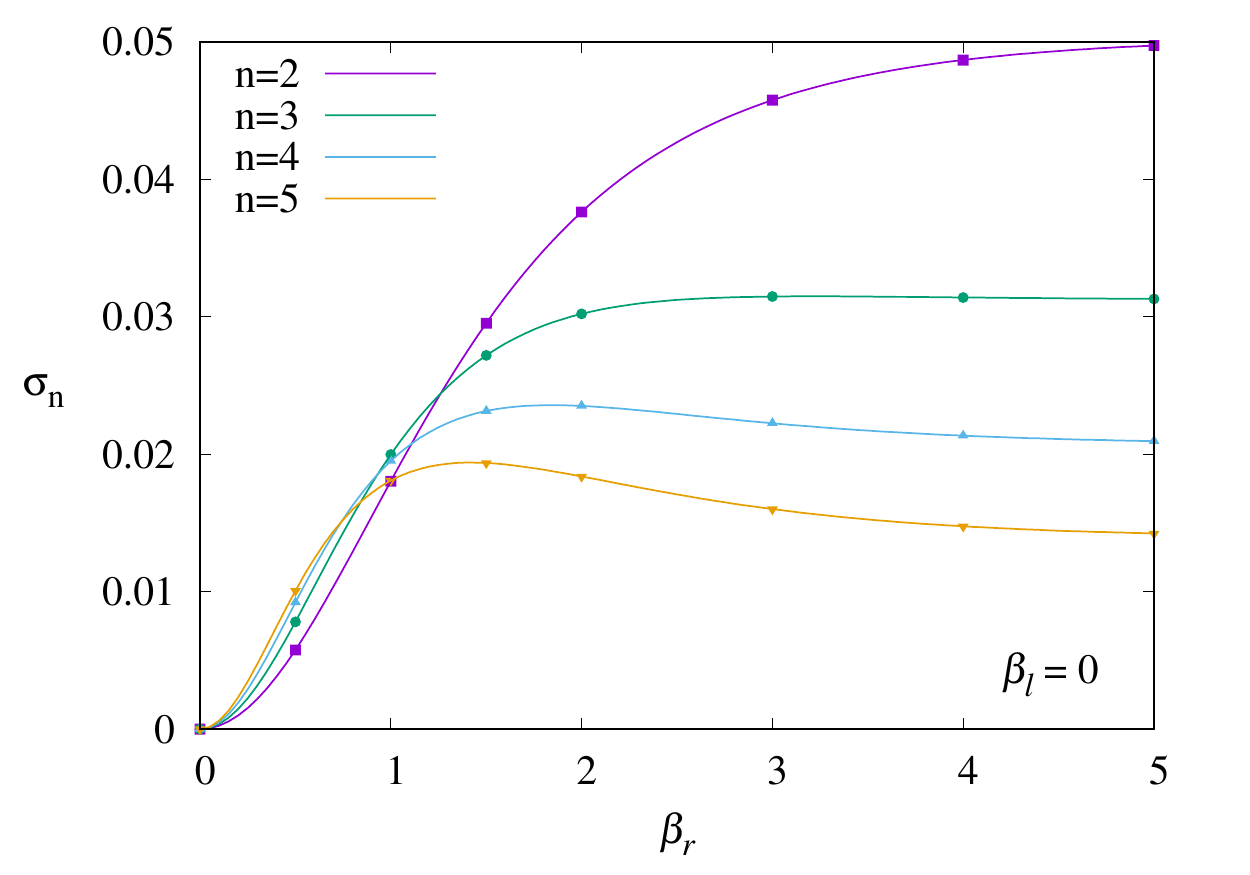}
\caption{Prefactor of the logarithm in the RMI of adjacent intervals,
calculated in the NESS $\rho_\infty$ for various $n$  (symbols), and compared to the analytic formula in Eq. \eqref{sign} (lines).}
\label{fig:sign}
\end{figure}

Before testing the validity of \eqref{sign}, let us first observe some of its
properties. First of all, due to the definition of the $\lambda_k$ in $\eqref{lamk}$,
the expression is real as it should. Moreover, it is easy to check that \eqref{sign}
is invariant under the simultaneous exchange of $a \to 1-a$ and $b \to 1-b$,
which follows from the symmetry of the R\'enyi entropy under a particle-hole
transformation. This property is necessary to account for the second jump at
$q\to\pm\pi$ of the actual occupation $n_q$ in \eqref{nq}. The actual scaling
of the RMI \eqref{rmi} in the NESS can then be obtained by calculating
\eq{
S_n(\rho_A) = \frac{1}{1-n} \ln \det \left[C^n_A + (1-C_A)^n\right] ,
}
using the matrix elements in \eqref{cmn} for increasing sizes $\ell$ of the
intervals $A_{1,2}$. We fixed $\beta_l=0$ and calculated the scaling of the RMI
for several values of $n$ by varying the inverse temperature $\beta_r$.
The fitted values of the logarithmic prefactor are shown in Fig.~\ref{fig:sign},
together with the analytical prediction \eqref{sign}, with an excellent agreement.

%
%

It is also straightforward to handle the case of disjoint intervals, where the symbol
\eqref{phi12} has four jumps, characterized by $\beta_1=\beta_3=\beta(\lambda_k)$
and $\beta_0=\beta_2=-\beta(\lambda_k)$.
One then has
\begin{align}
&\ln Z'_{A_1 \cup A_2,k} = 
\ell \ln \left[(1-a+\ee^{i\lambda_k} a)(1-b+\ee^{i\lambda_k}b)\right] 
\nonumber \\
&-2\beta^2(\lambda_k) \ln \left[\frac{f^2(\ell) f(2\ell+d)f(d)}{f^2(\ell+d)}\right]
+ 4 \, \mathcal{C}(\lambda_k) \, ,
\label{lnZ12}
\end{align}
where we introduced
\eq{
f(x) = L \sin ( \pi x/L) \, .
}
Thus in the limit $L\to\infty$ the RMI becomes
\eq{
\mathcal{I}_n = -\sigma_n \ln \eta,
\label{Indi}}
where the parameter $\eta$ is nothing but the cross ratio of the disjoint intervals
\eq{
\eta = \frac{(2\ell+d)d}{(\ell+d)^2}.
\label{cr}}
Note that this is exactly the result conjectured in \cite{AEF14}.

\begin{figure}[htb]
\center
\includegraphics[width=\columnwidth]{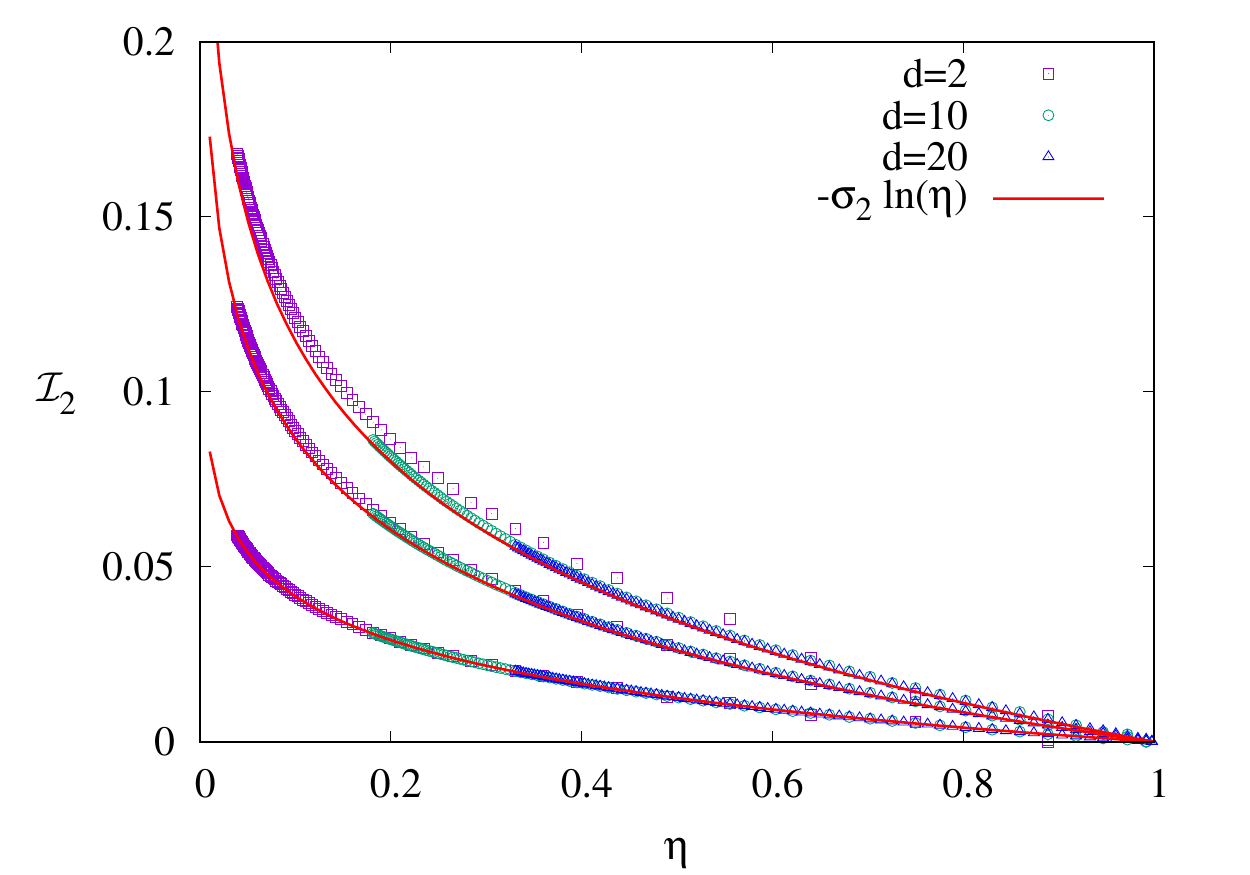}
\caption{RMI with $n=2$ for disjoint intervals in the NESS. The data is shown for fixed distances $d$,
varying $\ell$ and plotted as a function of the cross ratio $\eta$. The red solid lines show the
analytic result \eqref{Indi}. The inverse temperatures are $\beta_l=0$ and $\beta_r=5,2,1$
(from top to bottom).}
\label{fig:rmidi}
\end{figure}

The result can be checked against the numerical data obtained in the NESS,
by fixing the distance $d$ of the intervals and varying $\ell$. The RMI with $n=2$
is plotted against $\eta$ in Fig.~\ref{fig:rmidi} for $\beta_l=0$ and various $\beta_r$,
while the lines show the FH result in Eq. \eqref{Indi}. The agreement is good, although
some finite-size corrections can be seen for small values of $d$.

%
%

\subsection{Analytic continuation\label{sec:rmiac}}

Our next goal is to perform an analytic continuation of the RMI
to arbitrary R\'enyi index $\alpha$. We follow the route that has been applied in
\cite{FG22} to construct an integral representation of the sum \eqref{sign}.
Let us first note, that the discrete values \eqref{lamk} appearing in the sum
are actually related to the zeroes of the polynomial
\eq{
p_n(z) = z^n + (1-z)^n = \prod_{k=-\frac{n-1}{2}}^{\frac{n-1}{2}}
\left(1-\frac{z}{z_k}\right) .
\label{pnz}}
Indeed, it easy to see that
\eq{
z_k = (1-\ee^{i\lambda_k})^{-1},
\label{zk}}
and thus the summand of \eqref{sign} reads
\eq{
\ln^2 \left(\frac{1-a+\ee^{i\lambda_k} a}{1-b+\ee^{i\lambda_k} b}\right)=
\ln^2 \left(\frac{z_k-a}{z_k-b}\right) .
}
Analogously to \eqref{betalam}, we can define the function
\eq{
\beta(z) = \frac{1}{2\pi i}\ln \left(\frac{z-a}{z-b}\right) ,
\label{betaz}}
and by the residue theorem we thus have
\eq{
\sigma_n = 
-\frac{2}{1-n} \oint_\Gamma \frac{\dd z}{2\pi i}
\frac{p'_n(z)}{p_n(z)} \beta^2(z) \, ,
\label{signcint}}
where the integration contour $\Gamma$ encircles the roots $z_k$.

\begin{figure}[htb]
\center
\includegraphics[width=0.6\columnwidth]{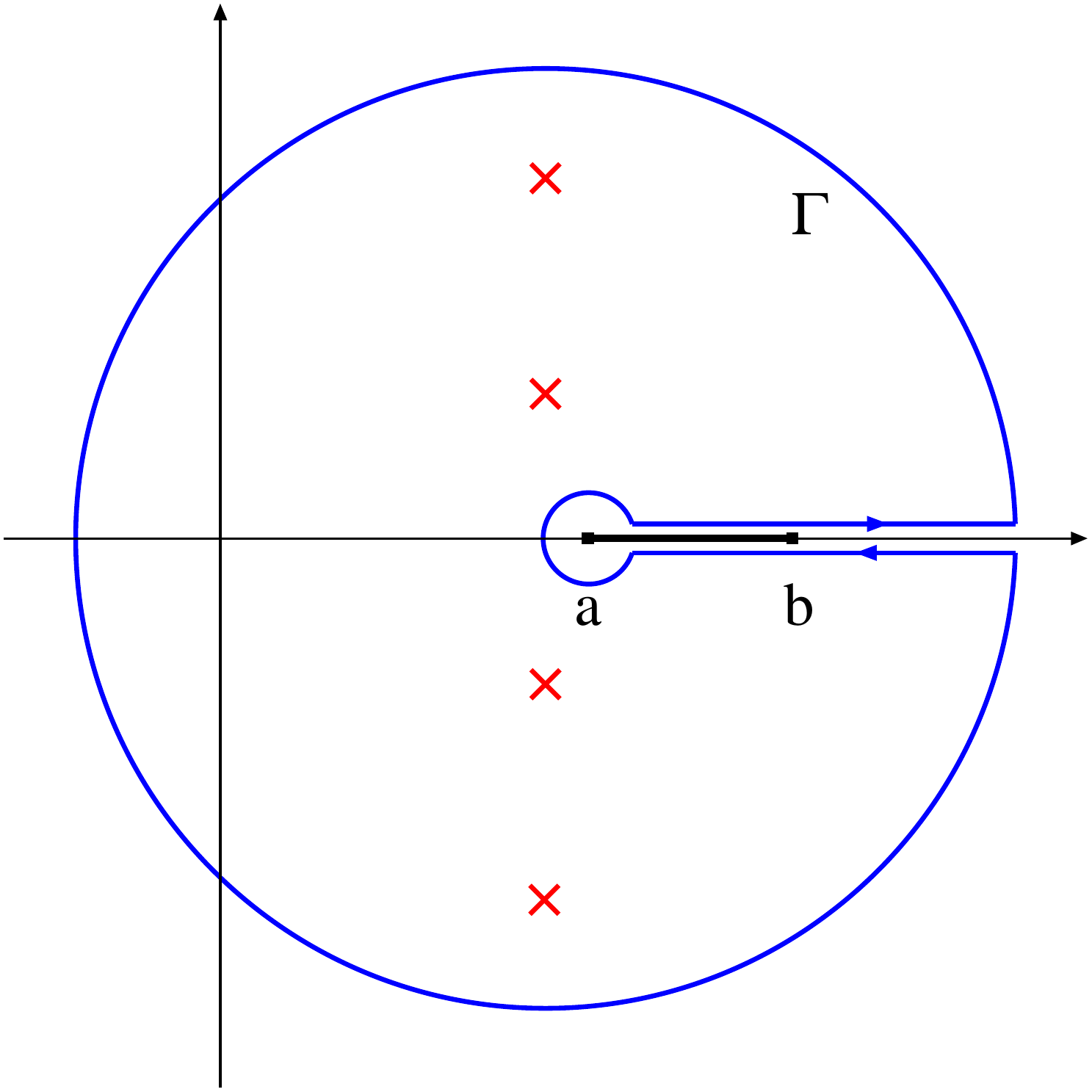}
\caption{Integration contour $\Gamma$ (blue) in \eqref{signcint}. The thick black line corresponds
to the branch cut, while the red crosses indicate the poles $z_k$ in \eqref{zk} for $n=4$.
}
\label{fig:cont}
\end{figure}

To evaluate the contour integral above, one should note that the function
$\beta(z)$ has a branch cut
\eq{
\beta(x+i0^\pm) = \frac{1}{2\pi i}\ln \left(\frac{x-a}{b-x}\right) \mp \frac{1}{2}
\label{betabc}}
along $x\in\left[a,b\right]$, while it is analytic outside of that interval.
The integration contour must avoid this branch cut, running infinitesimally
close to it on both sides, whereas the contour can be closed on a circle of radius $R \to \infty$,
as depicted in Fig.~\ref{fig:cont}.
Since $\beta(z) \to 0$ on the latter part of the contour, the only contribution to the
integral comes from the branch cut. Indeed, using \eqref{betabc}, one needs only the
term in $\beta^2(z)$ that changes sign  when crossing the branch cut.
Furthermore, one can rewrite
\eq{
\frac{1}{1-n}\frac{p'_n(x)}{p_n(x)} = s'_n(x)
}
as the derivative of the R\'enyi entropy density
\eq{
s_n(x) = \frac{1}{1-n} \ln \left[ x^n + (1-x)^n \right] .
}
This yields the integral representation of the prefactor
\eq{
\sigma_n = 
\frac{1}{\pi^2} \int_a^b  \dd x \, s'_n(x)
\ln \left(\frac{b-x}{x-a}\right).
\label{signint}}
The result is an analytic function of the index $n$ and
can thus be continued to arbitrary non-integer values.

%
%

It is instructive to check how \eqref{signint} relates to the result in
\cite{EZ14}, obtained by a direct FH calculation with the NESS correlation matrix \eqref{cmn},
using the method of \cite{JK04}. In fact, by partial integration one obtains
\begin{align}
\sigma_n
&=\frac{1}{\pi^2} \int \limits_{a+\epsilon}^{b-\epsilon} \dd x \, s_n(x) \frac{b-a}{(x-a)(b-x)} \nonumber \\
&+\frac{1}{\pi^2} \left[s_n(b)+s_n(a)\right] \ln \frac{\epsilon}{b-a} \, ,
\label{signpint}
\end{align}
which is exactly the expression found in \cite{EZ14} for $n=1$.
Note that one has to introduce an infinitesimal $\epsilon \to 0$
to avoid the logarithmic singularity of the integral, which is exactly canceled
by the boundary term on the second line of \eqref{signpint}.
The prefactor $\sigma=\lim_{n\to1}\sigma_n$ can be evaluated in terms of the
dilogarithm function $\Li{x}$ as \cite{EZ14}
\begin{align}
\sigma = \frac{1}{\pi^2} \left[
 a \, \Li{\frac{a-b}{a}} \right. + &(1-a) \Li{\frac{b-a}{1-a}} \nonumber \\
+ b \, \Li{\frac{b-a}{b}} + & \left. (1-b) \Li{\frac{a-b}{1-b}} \right] .
\end{align}

\begin{figure}[htb]
\center
\includegraphics[width=\columnwidth]{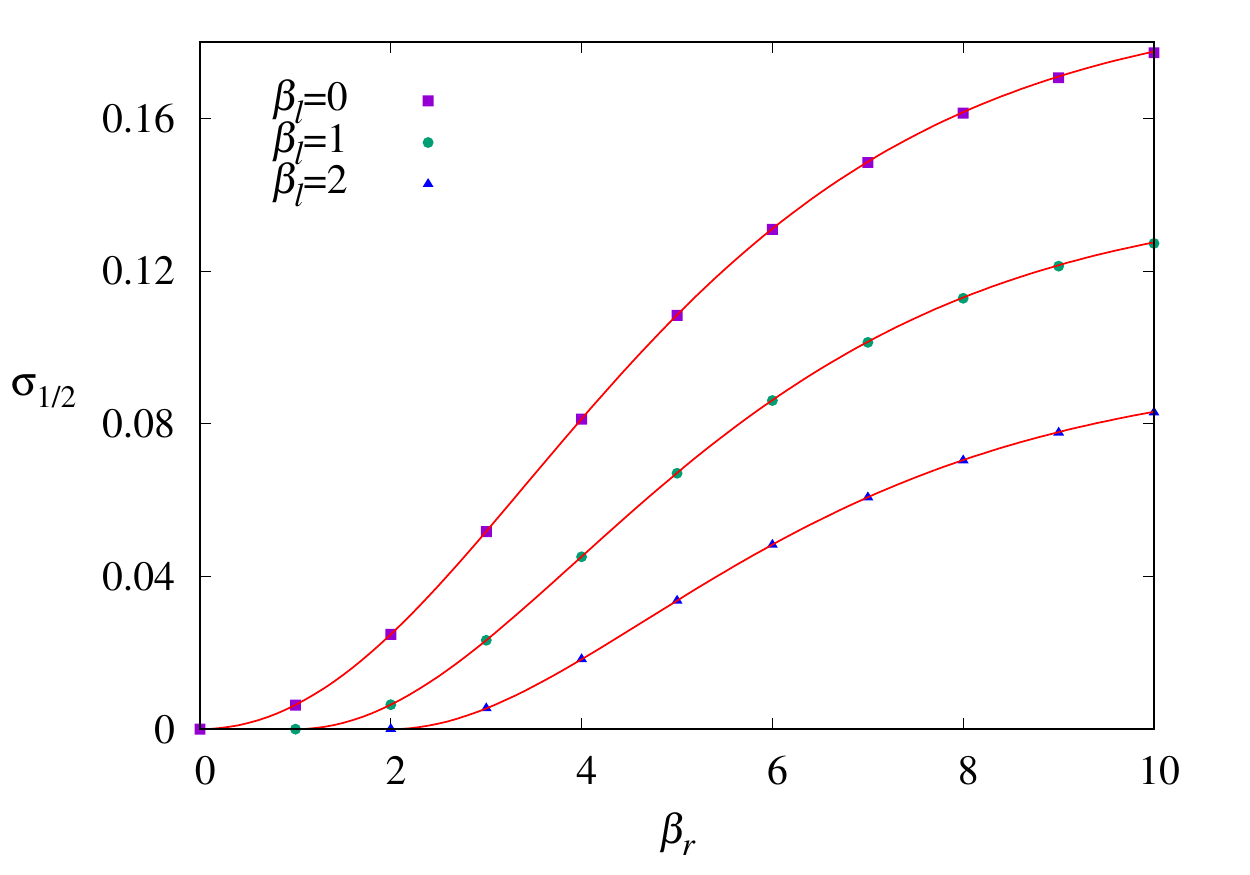}
\caption{Logarithmic prefactor of the RMI with $\alpha=1/2$, for various
pairs of $\beta_l$ and $\beta_r$. The symbols indicate the values obtained by data fits,
while the red solid lines show the analytical result \eqref{sig12}.}
\label{fig:sig12}
\end{figure}

Another case we are interested in is the RMI with index $\alpha=1/2$,
where the derivation of a closed form expression is more involved and
can be found in appendix \ref{app:sig}. The result can again be written
in terms of dilogarithms, but with complex arguments as
\begin{align}
\sigma_{1/2} &=
\frac{2}{\pi^2} \rp \left[\frac{1}{2}\ln^2 \left(\frac{1+i \, x_a}{1+i \, x_b}\right) 
\right. \nonumber \\ &\left.
+\Li{\frac{1+ i \, x^{\phantom{-1}}_a}{1+i \, x^{-1}_a}}+\Li{\frac{1+ i \, x^{\phantom{-1}}_b}{1+i \, x^{-1}_b}}
\right. \nonumber \\ &\left.
- \Li{\frac{1+ i \, x^{\phantom{-1}}_b}{1+i \, x^{-1}_a}}-\Li{\frac{1+ i \, x^{\phantom{-1}}_a}{1+i \, x^{-1}_b}}
\right],
\label{sig12}
\end{align}
where the parameters in the arguments are defined as
\eq{
x_a =\frac{\sqrt{1-a}-\sqrt{a}}{\sqrt{1-a}+\sqrt{a}} \, ,
\qquad
x_b = \frac{\sqrt{1-b}-\sqrt{b}}{\sqrt{1-b}+\sqrt{b}} \, .
\label{xab}}
The formula \eqref{sig12} is checked against the numerical data in Fig.~\ref{fig:sig12}
with an excellent agreement.

%
%

\section{Entanglement negativity\label{sec:neg}}

We now move forward to our main goal of evaluating the logarithmic
negativity between two intervals $A_1$ and $A_2$ in the NESS.
Throughout this section we will use the notation $A=A_1\cup A_2$,
and $\rho_A=\Trb (\rho_\infty)$ with $B = \bar A$.
The negativity is originally defined via the partial transpose of the
reduced density matrix $\rho_A$ which, however, in general
leads to a non-Gaussian state \cite{EZ15}. For free-fermion systems one could,
instead, define the negativity via the partial time reversal operation \cite{SSR17}.
This leads to the Gaussian operators
\eq{
O_+ = \rho_A^{T_2} \, , \qquad
O_- = (\rho_A^{T_2})^\dag \, ,
\label{Opm}}
where $T_2$ denotes the partial time reversal with respect to $A_2$.
They are characterized by their correlation matrices
\eq{
C_\pm =
\twomat{C_{A_1A_1}}{\pm i C_{A_1A_2}}
{\pm i C_{A_2A_1}}{\identity_{A_2} - C_{A_2A_2}},
\label{cpm}}
where we used a block notation and one has  $(C_\pm)_{mn}=\Tr(O_\pm c_m^\dag c_n)$.
The fermionic logarithmic negativity is then defined via the trace norm as \cite{SSR17}
\eq{
\lneg = \ln || \rho_A^{T_2}||_1 = \ln \Tr \sqrt{O_+ O_-} \; ,
\label{neg}}
and was shown to be a proper entanglement measure \cite{SR19}. In fact, it
also provides a natural upper bound to the standard negativity, as defined via
the partial transpose \cite{HW16,EEZ18}.

To calculate $\lneg$, we follow a similar strategy to the one applied for the RMI.
Namely, we use the replica trick and define the quantity
\eq{
\mathcal{E}_n = \ln \Tr (O_+ O_-)^{n/2}
\label{rneg}}
for arbitrary \emph{even} integer $n$. If we are able to calculate $\lneg_n$
analytically, the logarithmic negativity follows from the replica limit
\eq{
\lneg = \lim_{n \to1} \lneg_{n} \, .
\label{lnlim}}
One thus needs the analytic continuation $n\to1$ of a sequence
$\lneg_n$ on even integers. Analogously to the RMI in \eqref{sn},
this object can be decoupled into a sum \cite{SSR17}
\eq{
\mathcal{E}_n =  2 \, \rp \sum_{k=\frac{1}{2}}^{\frac{n-1}{2}} \ln \tilde Z_{k},
\label{rnegsum}}
where we defined the twisted partition function
\eq{
\tilde Z_{k}=
\Tr(\rho_A \ee^{i\lambda_k \hat N_{A_1}}  \ee^{i (\pi-\lambda_k) \hat N_{A_2}}) \, .
}
Here $\hat N_{A_1}$ and $\hat N_{A_2}$ are the particle-number operators of the respective
subsystem. Note that both $\lambda_k$ and the reversed phase $\pi-\lambda_k$
are located on the upper half plane for $k>0$, and take values as in \eqref{lamk} with half-integer $k$.

The partition function $\tilde Z_{k}$ can be computed using trace formulas for the products
of Gaussian operators as
\eq{
\tilde Z_{k}=\det \left[\identity_A -C_A + \ee^{i\lambda_k \identity_{A_1}}
\ee^{i (\pi-\lambda_k) \identity_{A_2}} C_A \right],
\label{detzt}}
where $\identity_{A_1}$, $\identity_{A_2}$ and $\identity_{A}$ are the identity matrices
on the respective subsystem. Clearly, due to the appearance of the reversed phase,
\eqref{detzt} is not the determinant of a Toeplitz matrix. We will show, however, that
after exchanging $C_A \to C'_A$, a dual Toeplitz matrix can again be constructed,
which is similar to the overlap matrix. To this end, we first note that for a correlation
matrix $C^0_A$ of a Fermi sea state as in \eqref{C0mn}, one has the identity
\eq{
\Tr(DC^0_A)^n = \Tr \tilde M^n \, ,
\label{trdc}}
where $D=\mathrm{diag}(d_j)$ is an arbitrary \emph{diagonal} matrix and
the modified overlap matrix is defined as
\eq{
\tilde M_{pq} =  \sum_{j \in A} 
d_j \, \varphi^*_p(j) \varphi^{\phantom{*}}_q(j) \, .
\label{tmpq}}
In other words, since \eqref{trdc} is satisfied for arbitrary $n$, the matrices
$DC^0_A$ and $\tilde M$ have the same eigenvalues. The proof of the identity
simply follows by writing out the left hand side of \eqref{trdc} as
\eq{
\sum_{j_1,\dots,j_n \in A} \sum_{q_1,\dots,q_n \le 0}
d_{j_1}   \varphi^*_{q_1}(j_1) \varphi^{\phantom{*}}_{q_1}(j_2) \dots
d_{j_n}   \varphi^*_{q_n}(j_n) \varphi^{\phantom{*}}_{q_n}(j_1)
}
and reordering the sums.

One can now replace $C_A \to C'_A$ as defined in \eqref{CpA}
and insert it into \eqref{detzt}, which leads to
\eq{
\tilde Z_{k}=
\det \left[\identity_A+DC^0_A\right] \det \left[E \right],
\label{detzt2}}
where the matrices are given by
\begin{align}
&D = E^{-1}(a-b)
(\ee^{i\lambda_k \identity_{A_1}} \ee^{i (\pi-\lambda_k) \identity_{A_2}}-\identity_A) \, , \\
&E = (1-b)\identity_A + b \, \ee^{i\lambda_k \identity_{A_1}} \ee^{i (\pi-\lambda_k) \identity_{A_2}}\, .
\end{align}
The matrix $D$ is thus diagonal, with matrix elements given by
\eq{
d_j =
\begin{cases}
\frac{(a-b)(\ee^{i\lambda_k}-1)}{1+b(\ee^{i\lambda_k}-1)} & j \in A_1 \\
\frac{(a-b)(\ee^{i (\pi-\lambda_k)}-1)}{1+b(\ee^{i (\pi-\lambda_k)}-1)} & j \in A_2 \\
\end{cases} .
}
Finally, we can apply \eqref{trdc} and replace $DC^0_A$ by $\tilde M$ in the
first determinant of \eqref{detzt2}. In the limit $L\to \infty$, this becomes a Toeplitz
determinant of the matrix
\eq{
(\identity + \tilde M)_{pq} = \int_{0}^{2\pi} \tilde \phi_A(\theta) \ee^{-i(p-q)\theta} \dd \theta \, ,
\label{Mtoep}}
with the Fisher-Hartwig symbol
\eq{
\tilde \phi_A(\theta) = \begin{cases}
\Phi(\lambda_k) & 0<\theta< \theta_1 \\
\Phi(\pi-\lambda_k) & \theta_2 < \theta < \theta_3 \\
1 & \mathrm{otherwise}
\end{cases},
\label{phitk}}
where $\Phi(\lambda)$ is given in \eqref{Phi}, while 
the jump locations $\theta_1$ and $\theta_2,\theta_3$ are defined in
\eqref{theta1} and \eqref{theta23}, respectively.

We are now ready to evaluate the determinant using the FH theorem \eqref{FH}.
Let us first consider the case of adjacent intervals where $\theta_2=\theta_1$
and $\theta_3=2\theta_1$. One has thus three FH singularities described
by the functions
\eq{
\begin{split}
&\beta_{0}(\lambda_k) = -\frac{1}{2\pi i} 
\ln [\Phi(\lambda_k)] \, , \\
&\beta_{1}(\lambda_k) = \frac{1}{2\pi i} 
\ln \left[\frac{\Phi(\lambda_k)}{\Phi(\pi-\lambda_k)}\right],  \\
&\beta_{2}(\lambda_k) = \frac{1}{2\pi i} 
\ln [\Phi(\pi-\lambda_k)] \, .
\end{split}
\label{betaneg}}
In turn, the R\'enyi negativity \eqref{rnegsum} is given by
\eq{
\lneg_n = \tilde \alpha_n \, \ell + \tilde\sigma_n \ln \ell + \const,
}
where the prefactor of the logarithmic contribution is
\eq{
\tilde\sigma_n = 4 \, \rp \sum_{k=\frac{1}{2}}^{\frac{n-1}{2}}
\sum_{s_1<s_2}  \beta_{s_1}(\lambda_k) \beta_{s_2}(\lambda_k) \, .
}
Note that, in general, one has an extensive term which receives
contributions also from the second determinant in \eqref{detzt2},
and its prefactor can be evaluated as
\eq{
\tilde \alpha_n =2 \, \rp \sum_{k=\frac{1}{2}}^{\frac{n-1}{2}}
\ln \left[(1-a+\ee^{i\lambda_k} a)(1-b+\ee^{i\lambda_k}b)\right] .
}
Here we used the fact that $\pi-\lambda_k$ has the same set of allowed
values as $\lambda_k$. Furthermore, taking twice the real part is equivalent
to including the negative $k$ values in the sum. Using the factorization in
\eqref{pnz}, one has then
\eq{
\tilde \alpha_n = 
\ln\left[a^n + (1-a)^{n}\right] +
\ln\left[b^n + (1-b)^{n}\right],
}
and hence the extensive prefactor vanishes in the limit $n \to 1$.
On the other hand, the logarithmic prefactor can be obtained
using \eqref{betaneg} as
\eq{
\tilde\sigma_n =
\frac{1}{\pi^2} \rp \sum_{k=\frac{1}{2}}^{\frac{n-1}{2}} \left\{
2\ln^2 [\Phi(\lambda_k)]
-\ln [\Phi(\lambda_k)] \ln [\Phi(\pi-\lambda_k)]
\right\}.
\label{signegn}}

Finally, let us consider the case of disjoint intervals. The symbol \eqref{phitk}
has then four FH singularities, characterized by
\eq{
\begin{split}
&\beta_{1}(\lambda_k) = -\beta_{0}(\lambda_k) =
\frac{1}{2\pi i}  \ln [\Phi(\lambda_k)] \, , \\
&\beta_{3}(\lambda_k) = -\beta_{2}(\lambda_k) =
\frac{1}{2\pi i}  \ln [\Phi(\pi-\lambda_k)] \, .
\end{split}
\label{betaneg2}}
Similarly to \eqref{lnZ12}, the different pairings of the $\beta_s(\lambda_k)$
are now multiplied by different logarithmic factors, and the prefactor \eqref{signegn}
splits into two parts. In particular, the first sum including the $\ln^2[\Phi(\lambda_k)]$
is multiplied by $\ln \ell$, whereas the second mixed term is multiplied by $-\ln \eta$,
with the cross-ratio defined in \eqref{cr}.

\subsection{Analytic continuation}

The final step to obtain the entanglement negativity is to carry out the
analytic continuation \eqref{lnlim}. This can be performed in a very similar
fashion as for the R\'enyi mutual information in Sec. \ref{sec:rmiac}.
Namely, we shall find an integral representation of the sum \eqref{signegn},
which is analytic in $n$ and thus the limit $n \to 1$ can be carried out.
First of all note that, using \eqref{Phi}, the first part of the sum is actually related
to the RMI prefactor \eqref{sign} as
\eq{
\frac{2}{\pi^2} \rp \sum_{k=\frac{1}{2}}^{\frac{n-1}{2}}
\ln^2 [\Phi(\lambda_k)] = 2(1-n)\sigma_n \, ,
}
and thus will vanish in the limit $n \to 1$. We shall thus focus only on the mixed
term in \eqref{signegn}, where both $\lambda_k$ and $\pi-\lambda_k$ appear.

The main idea that was already applied in \cite{FG22} is to consider the complex
roots $\tilde z_k$ of the polynomial
\eq{
\tilde p_n(z) = z^{n/2} + (1-z)^{n/2} = 
\prod_{k=\frac{1}{2}}^{\frac{n-1}{2}} \left(1-\frac{z}{\tilde z_k} \right),
\label{tpn}}
which can be expressed via the positive phases $\lambda_k$ as
\eq{
\tilde z_k^{-1} = 1+\ee^{-2i\lambda_k} \, , \qquad
k = \frac{1}{2}, \dots, \frac{n-1}{2} \, .
\label{ztk}}
In the integral representation the roots should appear as poles.
It should be noted, however, that there is a one-to-one correspondence between the
phases $\lambda_k$ and the roots $\tilde z_k$ only for $n/2$ even.
Indeed, for $n/2$ odd the phase $\lambda_k=\pi/2$ appears, which
corresponds to $\tilde z_k^{-1}=0$, i.e. $\tilde p_n(z)$ actually has only $n/2-1$ roots.
The contribution of the missing pole thus has to be added in this case.
For the moment we shall assume $n/2$ to be even, and comment on the
other case later on.

Inverting the relation \eqref{ztk} requires some care. In order to have all the
phases on the upper half plane, one needs
\eq{
\ee^{i\lambda_k} = i\sqrt{\frac{\tilde z_k}{\tilde z_k-1}} ,
\qquad
\ee^{i(\pi-\lambda_k)} = i\sqrt{\frac{\tilde z_k-1}{\tilde z_k}} .
}
The integral representation of the sum thus reads
\eq{
\frac{1}{\pi^2} \mathrm{Re} 
\int_{\tilde \Gamma} \frac{\dd z}{2\pi i} \frac{\tilde p'_n(z)}{\tilde p_n(z)}
\ln [\Phi(z)]
\ln [\bar \Phi(z)] \, ,
\label{cintneg}}
where the arguments are given by
\begin{align}
&\Phi(z)= \frac{1-a+ i\sqrt{\frac{z}{z-1}} a}{1-b+ i\sqrt{\frac{z}{z-1}} b} \, , \\
&\bar\Phi(z)= \frac{1-b+ i\sqrt{\frac{z-1}{z}} b}{1-a+ i\sqrt{\frac{z-1}{z}} a} \, .
\end{align}
Note that, apart from the poles at $\tilde z_k$, the integrand has a branch cut
along $\left[0,1\right]$ due to the square root in the argument of the logarithms.
Thus the contour $\tilde \Gamma$ has to be chosen in a similar fashion to Fig.~\ref{fig:cont},
such that it goes around the branch cut and encircles all the poles, with its outer radius taken
to be very large.

The first contribution to the contour integral comes from the path running
along the branch cut. Indeed, for $z=x+i 0^\pm$ and $x \in \left[0,1\right]$ one has
\eq{
i\sqrt{\frac{z}{z-1}} \to \pm \sqrt{\frac{x}{1-x}} \, ,
\quad
i\sqrt{\frac{z-1}{z}} \to \mp \sqrt{\frac{1-x}{x}} \, ,
}
and the integral reads
\eq{
\frac{1}{\pi^2} \mathrm{Im} 
\int_{0}^{1} \frac{\dd x}{\pi} \frac{\tilde p'_n(x)}{\tilde p_n(x)}
\ln [\Phi(x)]
\ln [\bar \Phi(x)] \, ,
\label{intneg}}
where
\begin{align}
&\Phi(x)= \frac{1-a+ \sqrt{\frac{x}{1-x}} a}{1-b+ \sqrt{\frac{x}{1-x}} b} \, , \\
&\bar\Phi(x)= \frac{1-b- \sqrt{\frac{1-x}{x}} b}{1-a - \sqrt{\frac{1-x}{x}} a} \, .
\end{align}
Note that a factor two appears since the contributions are equal on both sides of the branch cut,
and by dropping the factor $i$ one now needs the imaginary part of the integral.
Since $\Phi(x)>0$ for $x\in[0,1]$, an imaginary part can only appear when $\bar \Phi(x)<0$.
This is the case on the interval $x \in [\tilde a , \tilde b ]$ where
\eq{
\tilde a = \frac{a^2}{a^2+(1-a)^2} \, ,
\qquad
\tilde b = \frac{b^2}{b^2+(1-b)^2} \, .
}
One has then $\ip \ln [\bar \Phi(x)]= \pi$ and one arrives at
\eq{
\frac{1}{\pi^2}
\int_{\tilde a}^{\tilde b} \dd x \, \frac{\tilde p'_n(x)}{\tilde p_n(x)}
\ln [\Phi(x)] \, .
}

We also need the contribution over the large circle of the contour $\tilde \Gamma$,
parametrized as $z=R \ee^{i\varphi}$ with $R\to\infty$, which turns out to be nonvanishing.
Indeed, the square-roots in $\Phi(z)$ and $\bar\Phi(z)$ converge to one and thus the
logarithms in the integrand yield a constant. On the other hand on has
\eq{
\lim_{R\to\infty} \frac{\tilde p'_n(z)}{\tilde p_n(z)}=
\begin{cases}
\frac{n}{2} \, z^{-1} & \textrm{$n/2$ even} \\
(\frac{n}{2}-1) \, z^{-1} & \textrm{$n/2$ odd}
\end{cases},
}
and thus the asymptotics depends on the parity of $n/2$. In fact, this is simply due to the
missing root for $n/2$ odd, where the contribution from $\lambda_k=\pi/2$ has to be added
to the contour integral \eqref{cintneg}. Adding the two pieces removes the parity dependence,
and with $\dd z/z=i \dd\varphi$ one obtains the contribution on the circle
\eq{
-\frac{n}{2\pi^2} \mathrm{Re} \ln^2 \left(\frac{1-a + a \, i}{1-b + b \, i} \right) ,
}
which is valid for arbitrary $n$.

Collecting all the contributions, the integral representation of the sum \eqref{signegn} reads
\begin{align}
\tilde \sigma_n &=
\frac{1}{\pi^2}
\int_{\tilde a}^{\tilde b} \dd x \frac{\tilde p'_n(x)}{\tilde p_n(x)}
\ln \left(\frac{1-a + a \, \sqrt{\frac{x}{1-x}}}{1-b + b \, \sqrt{\frac{x}{1-x}}}\right)
\nonumber \\
&- \frac{n}{2\pi^2} \, \rp \ln^2 \left(\frac{1-a + a \, i}{1-b + b \, i} \right)
+2(1-n)\sigma_n \, .
\label{signegint}
\end{align}
The result can now be analytically continued to $n\to 1$.
In fact, as shown in Appendix \ref{app:sig}, the integral for
$\tilde \sigma = \lim_{n\to1} \tilde \sigma_n$ can be evaluated
in a closed form. After a lengthy calculation one obtains
\begin{align}
\tilde \sigma &= \frac{1}{\pi^2}\rp \left[
\frac{1}{2}\ln^2 \left(\frac{1+i \, \tilde x_a}{1+i \, \tilde x_b}\right)
\right. \nonumber \\ & \left.
+ \Li{\frac{1+ i \, \tilde x^{\phantom{-1}}_b}{1-i \, \tilde x^{-1}_a}}
+ \Li{\frac{1+ i \, \tilde x^{\phantom{-1}}_a}{1-i \, \tilde x^{-1}_b}}
\right. \nonumber  \\ & \left.
- \Li{\frac{1+ i \, \tilde x^{\phantom{-1}}_a}{1-i \, \tilde x^{-1}_a}}
- \Li{\frac{1+ i \, \tilde x^{\phantom{-1}}_b}{1-i \, \tilde x^{-1}_b}} 
\right] ,
\label{signeg}
\end{align}
where the parameters are defined as
\eq{
\tilde x_a = 1-2a \, , \qquad \tilde x_b = 1-2b \, .
\label{txab}}

\subsection{Numerical results}

\begin{figure}[htb]
\center
\includegraphics[width=\columnwidth]{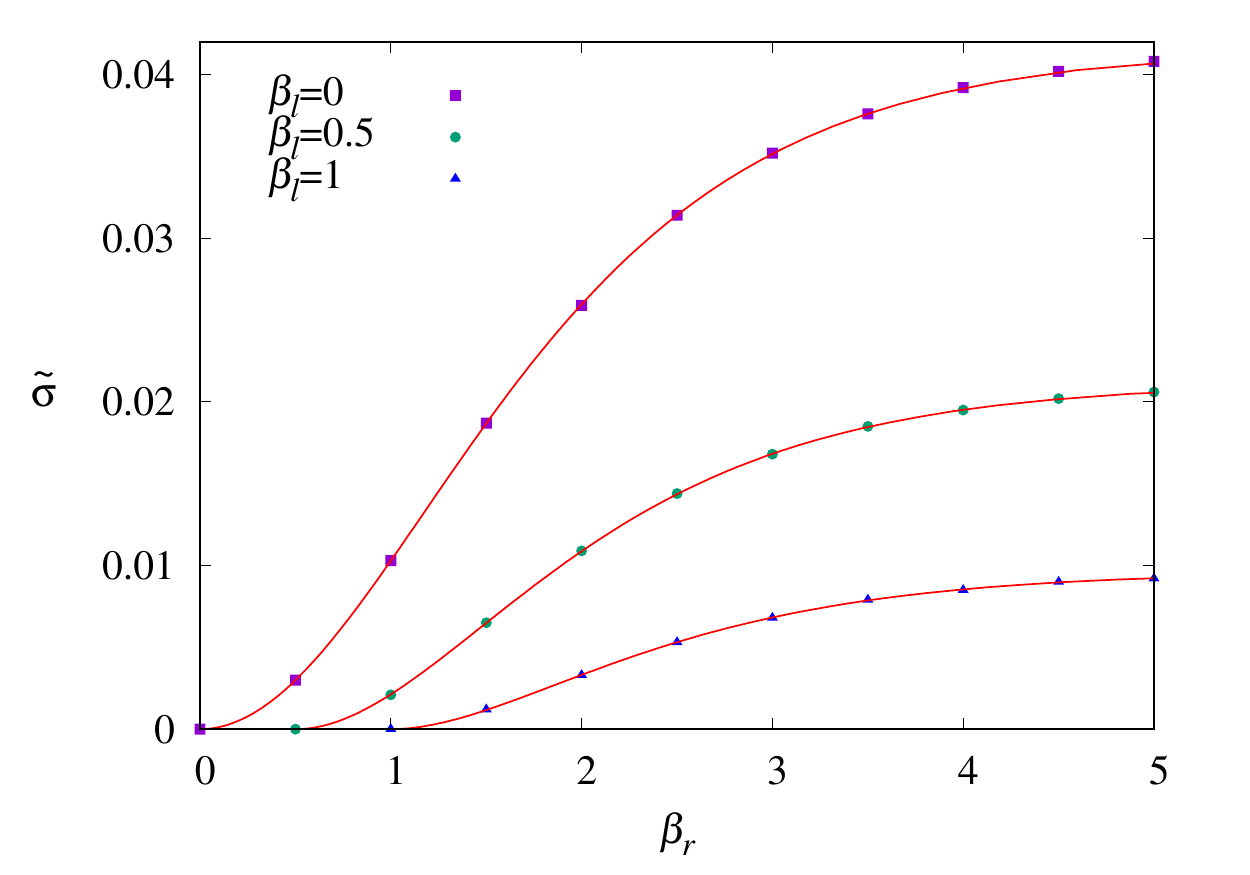}
\caption{Prefactor of the logarithmic scaling in the entanglement negativity for various
pairs of $\beta_l$ and $\beta_r$. The symbols indicate the values obtained by data fits,
while the red solid lines show the analytical result \eqref{signeg}.}
\label{fig:signeg}
\end{figure}

In the following we shall test the analytical results of the previous section
by calculating the entanglement negativity $\lneg=\ln \Tr \sqrt{O_+O_-}$ in the NESS. 
Using trace formulas for the product of Gaussian operators \cite{FC10}, this can be obtained
directly as \cite{SSR17,EEZ18}
\begin{align}
\lneg = &\ln \det \left[
\sqrt{\frac{\identity+G_\times}{2}} +
\sqrt{\frac{\identity-G_\times}{2}} \right] \nonumber \\
+ \frac{1}{2} &\ln \det \left[\frac{\identity+G_+G_-}{2}\right] ,
\label{lndet}
\end{align}
where $\identity\equiv\identity_A$ and we introduced the matrix
\eq{
G_\times =
\identity - (1-G_-)(\identity + G_+G_-)^{-1}(1-G_+) \, ,
\label{Gx}}
and $G_\pm = 2C_\pm -\identity$ with the correlation matrices
$C_\pm$ defined in \eqref{cpm}. Thus one has to evaluate determinants
of matrices that are completely determined via the reduced correlation matrix $C_A$.
We performed numerical calculations using \eqref{lndet} for various
pairs of inverse temperatures. For adjacent intervals
the analytical results of the previous section suggest the logarithmic scaling
\eq{
\lneg = \tilde \sigma \ln \ell + \const,
}
which we indeed observe in the numerics. We thus fitted our numerical data
and compared the prefactor $\tilde \sigma$ to the analytical formula \eqref{signeg},
with the result shown in Fig.~\ref{fig:signeg}. The agreement is excellent,
confirming that the logarithmic term in the negativity is determined only by the
values $a$ and $b$ on both sides of the jump in the occupation function.

%
%

\begin{figure}[htb]
\center
\includegraphics[width=\columnwidth]{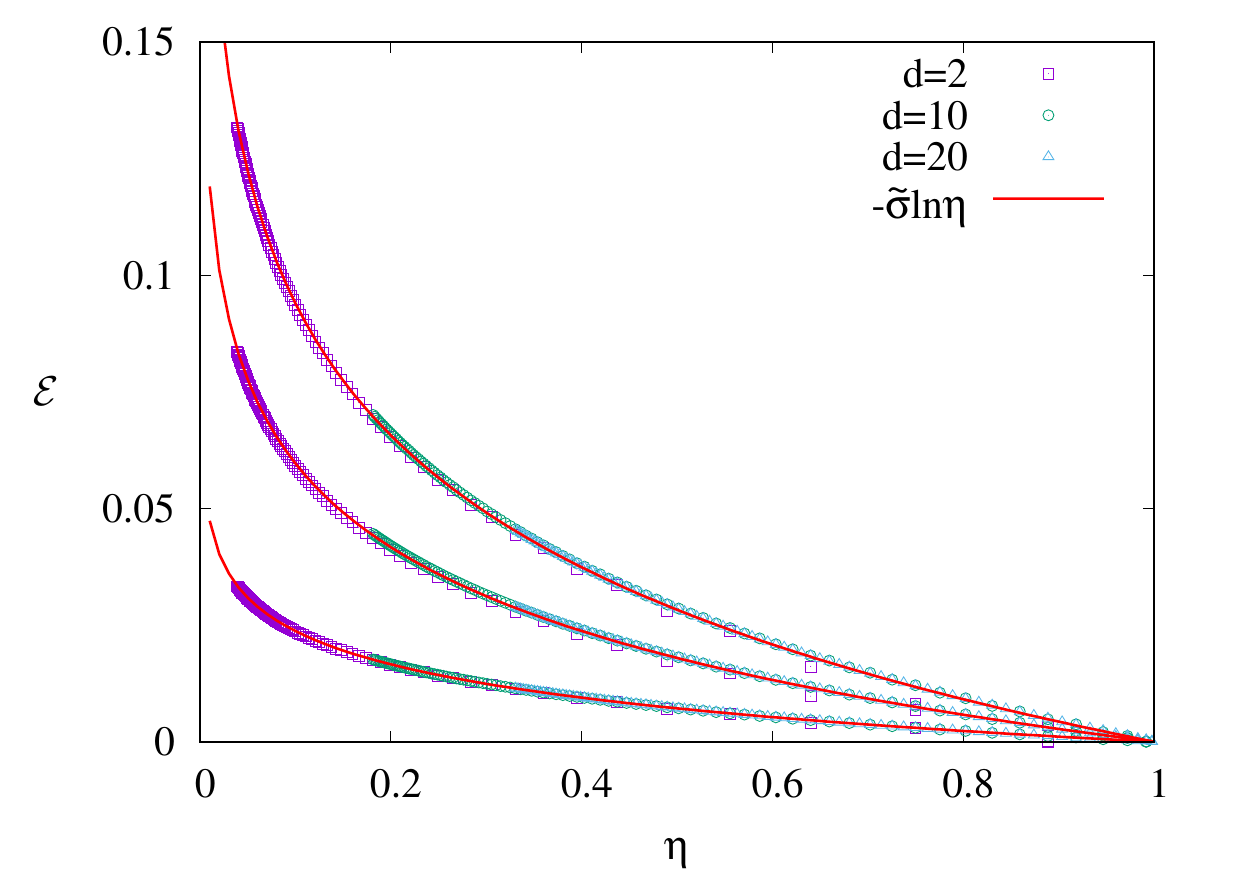}
\caption{Negativity for disjoint intervals in the NESS, plotted against the cross ratio $\eta$.
The parameters are $\beta_l=0$ and $\beta_r=5,2,1$ (from top to bottom)}
\label{fig:negdi}
\end{figure}

We have also checked the disjoint case, where the analytic continuation suggests the result
\eq{
\lneg = -\tilde \sigma \ln \eta \, ,
\label{lndi}}
in terms of the cross ratio \eqref{cr}. Note that, analogously to the result \eqref{Indi} for
the RMI, we expect that the constant term vanishes for disjoint intervals. This is indeed
what we observe, as demonstrated in Fig.~\ref{fig:negdi} for fixed $\beta_l=0$ and various $\beta_r$ values.
Remarkably, compared to the case of the RMI in Fig.~\ref{fig:rmidi}, the corrections
to \eqref{lndi} remain very small even for short distances $d=2$.

%
%

\subsection{Comparison of $\tilde \sigma$ and $\sigma_{1/2}$}

We are now ready to compare the RMI with $\alpha=1/2$ to the
entanglement negativity. In fact, previous results on quantum quenches in free-fermion
systems have indicated, that the two quantities are related as
$\lneg = \frac{1}{2}\mathcal{I}_{1/2}$ \cite{AC19,BKL22}. In the present case, however,
one can see that an analogous relation between the prefactors $\tilde \sigma$ and $\frac{1}{2}\sigma_{1/2}$
does not hold. Despite the very similar structure of their analytical expressions in \eqref{sig12} and \eqref{signeg},
the first obvious difference is in their respective variables \eqref{xab} and \eqref{txab}.
It is instructive to rewrite these variables in terms of the inverse temperatures. Substituting \eqref{ab} one obtains
\eq{
x_a= - \tanh \left(\frac{\beta_r}{4}\right), \qquad
\tilde x_a= - \tanh\left(\frac{\beta_r}{2}\right),
\label{xbeta}}
and analogously for $x_b$ and $\tilde x_b$ by exchanging $\beta_r \to \beta_l$.
Interestingly, there is a simple factor two difference, which might suggest that
the two prefactors become comparable only after an additional rescaling of the temperatures.
This is indeed how the pairs $\beta_l$ and $\beta_r$ were chosen in Figs.~\ref{fig:sig12}
and \ref{fig:signeg}, resulting in a qualitatively similar behaviour of the prefactors.
However, having a closer look at the vertical scale, one immediately sees that
the values of $\frac{1}{2}\sigma_{1/2}$ are much larger than those of $\tilde \sigma$.
The reason of this mismatch is that the arguments of the dilogarithm functions
\eqref{sig12} and \eqref{signeg} are still not the same in terms of their natural variables.

\begin{figure}[htb]
\center
\includegraphics[width=\columnwidth]{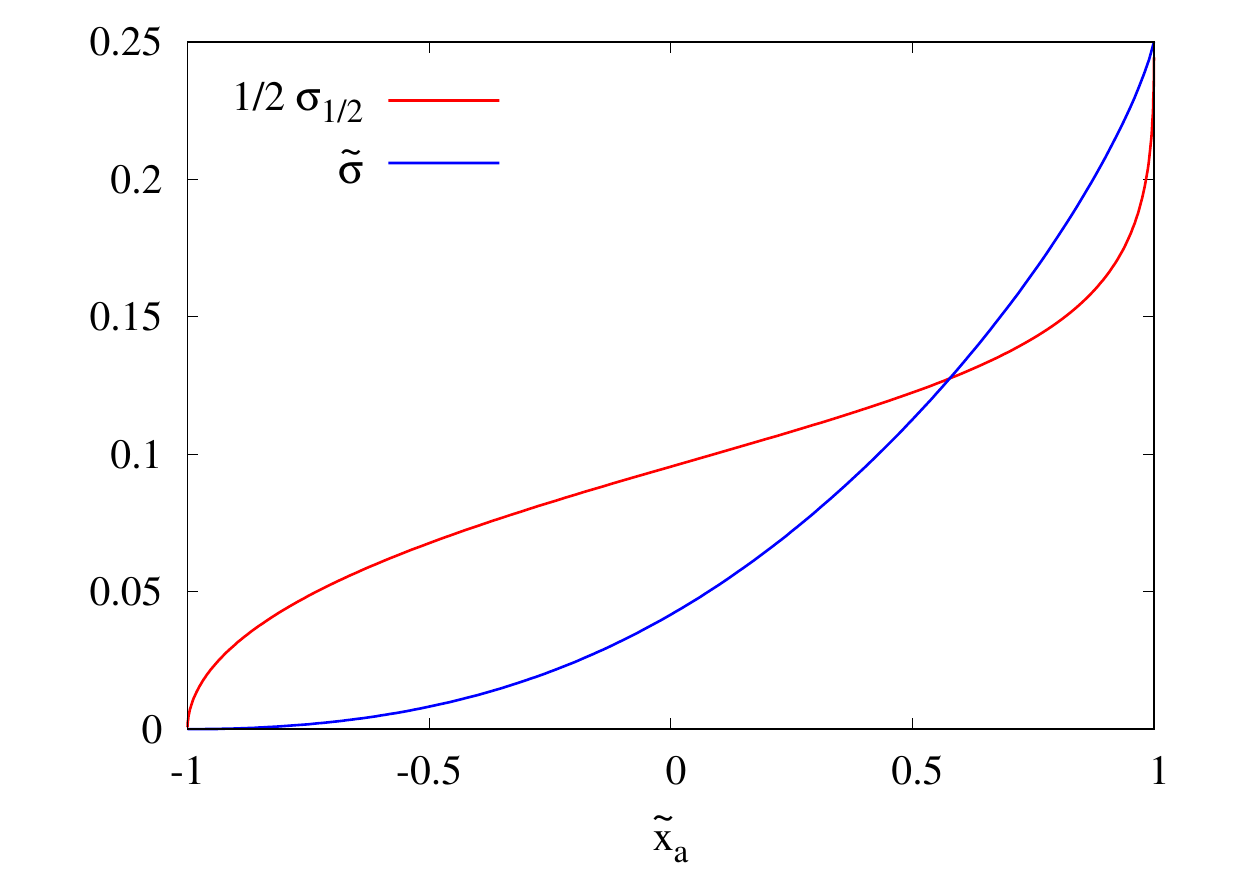}
\caption{Comparison of the $\alpha=1/2$ RMI and negativity prefactors with paramaters $x_b=\tilde x_b=-1$,
as a function of $\tilde x_a$.}
\label{fig:negvsrmi}
\end{figure}

In order to compare the prefactors directly, we shall fix $x_b=\tilde x_b=-1$ (setting
the left reservoir to zero temperature $\beta_l \to \infty$), and plot them against the
variable $\tilde x_a$, as shown in Fig.~\ref{fig:negvsrmi}. 
In the regime $\tilde x_a<0$ allowed for positive temperatures, $\frac{1}{2}\sigma_{1/2}$ is always larger
than $\tilde \sigma$, as observed already in Figs.~\ref{fig:sig12} and \ref{fig:signeg}.
 We have, however, also plotted the prefactors for $\tilde x_a>0$,
corresponding to negative temperatures according to \eqref{xbeta}. This should be understood as an
additional particle-hole transformation in the right bath, allowing for values $0<a<1/2$ in \eqref{nqpc}.
Interestingly, the two curves cross each other at $\tilde x_a \approx 0.577$, and $\tilde \sigma$ becomes larger than $\frac{1}{2}\sigma_{1/2}$.
The prefactors coincide again in the limit $\tilde x_a=1$, where the occupation \eqref{nqpc} with $a=0$ and $b=1$
is just a shifted Fermi sea. Hence, in this limit one reproduces the ground-state result
$\tilde \sigma=\frac{1}{2}\sigma_{1/2}=\frac{1}{4}$, which follows also from CFT calculations \cite{CCT12,CCT13}.

%
%

\subsection{Time evolution of the negativity}

Finally we investigate how the negativity builds up during the time
evolution leading to the NESS. In particular, we are interested in $\lneg(t)$
evaluated in the time-evolved state \eqref{rhot}. Since the density matrix $\rho_t$ is Gaussian
at any time $t$, the calculation of the negativity can easily be generalized. In fact, one
only needs the time-evolved correlation matrix
\eq{
C(t) = U^\dag C(0) U,
\label{Ct}}
where the propagator has matrix elements $U_{mn}=i^{n-m}J_{n-m}(t)$
given by the Bessel functions. The initial correlation matrix $C(0)=C_l(0) \oplus C_r(0)$ 
is a direct sum of thermal correlation matrices of the half-chains, corresponding to the factorized
initial state \eqref{rho0}. Note that, although in \eqref{Ct} one needs the product of infinite matrices,
the terms that contribute are heavily restricted by the light-cone behaviour of the propagator.

%
\begin{figure*}[htb]
\center
\includegraphics[width=\columnwidth]{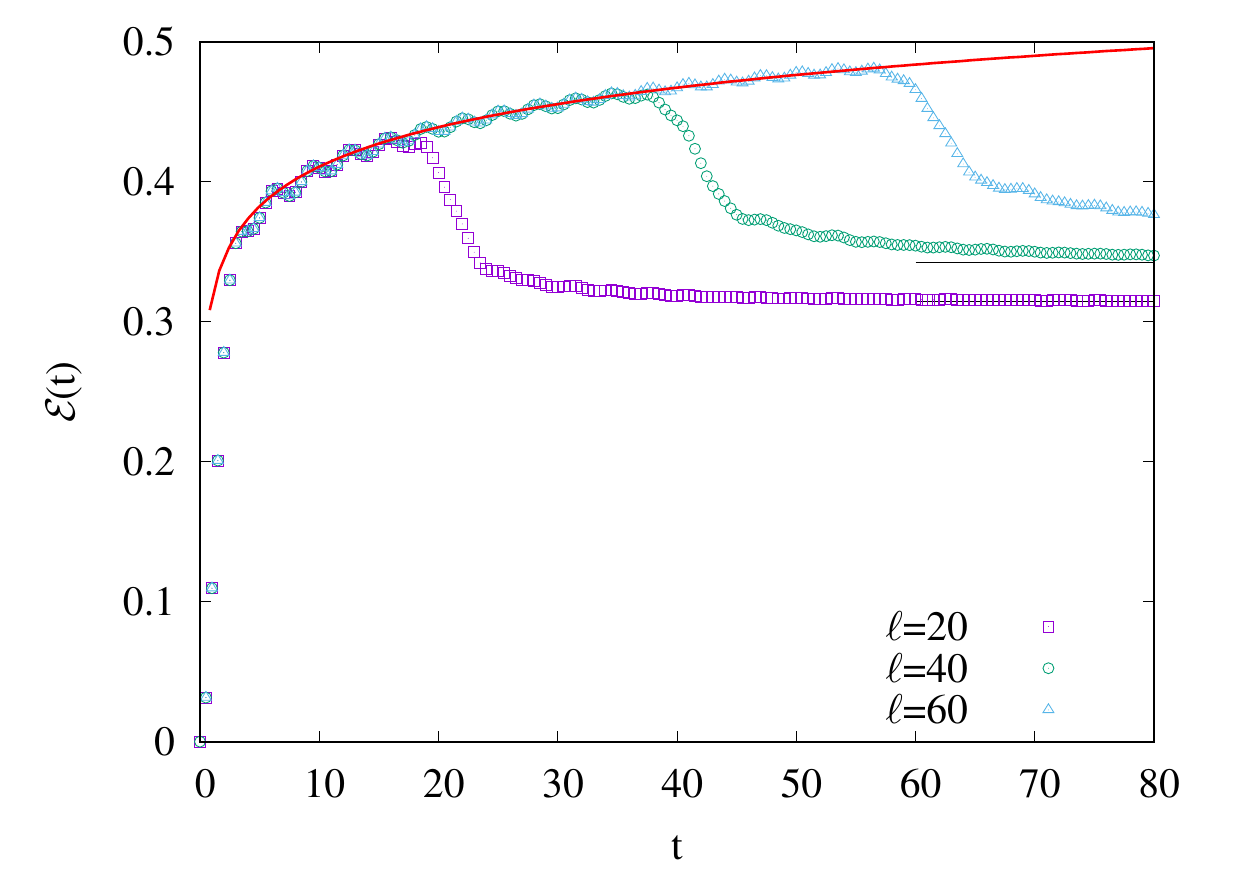}
\includegraphics[width=\columnwidth]{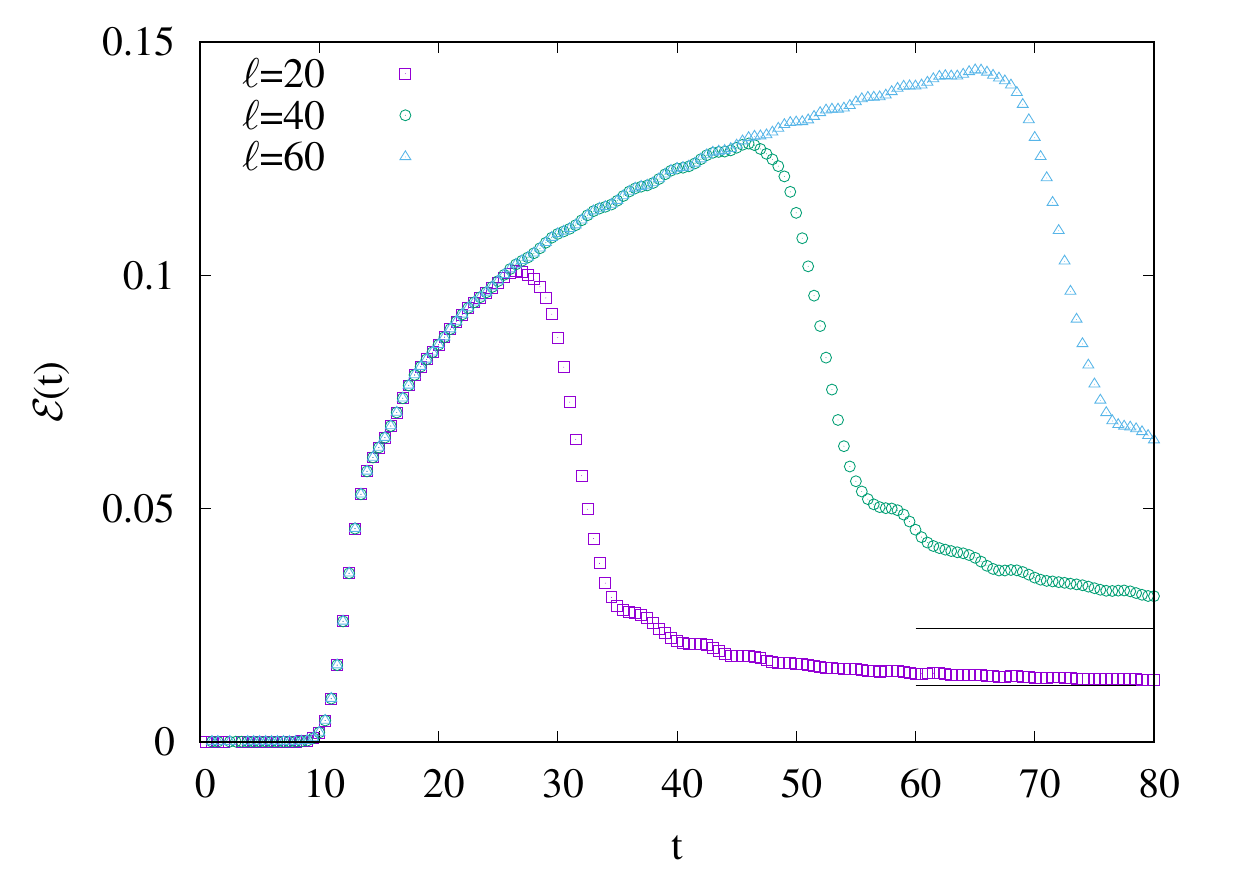}
\caption{Time evolution of entanglement negativity after connecting two half-chains at
inverse temperatures $\beta_l=0$ and $\beta_r$=5, for various subsystems sizes $\ell$.
Left: adjacent intervals. The red solid line shows the ansatz \eqref{lnegt},
while horizontal lines indicate the NESS results. Right: disjoint intervals with $d=20$.}
\label{fig:lnegt}
\end{figure*}
%

With the expression \eqref{Ct} at hand, one can simply define the time-dependent
matrices $G_\pm(t)$ and $G_\times(t)$ and use the corresponding formula \eqref{lndet}
to evaluate $\lneg(t)$. We are interested in the entanglement negativity of two intervals
placed symmetrically around the junction. The results are shown on Fig.~\ref{fig:lnegt}
for some fixed initial temperatures, both for adjacent as well as disjoint intervals.
Although the lack of translational invariance prevents us from using FH techniques,
we expect that the result is similar to the one found for the mutual information \cite{EZ14}.
Namely for adjacent intervals one has a logarithmic growth
\eq{
\lneg(t)=\tilde \sigma \ln t + \const
\label{lnegt}}
for short times $t < \ell$, with the exact same prefactor that governs the system-size scaling in the NESS.
On the other hand, for times $t>\ell$ after the front has crossed the boundaries of the intervals,
one expects a relaxation towards the NESS value.

This is indeed what we observe on the left of Fig.~\ref{fig:lnegt}, where \eqref{lnegt} is shown
by the red solid line, with the constant parameter determined by fitting. The NESS values are
indicated by the horizontal lines for $\ell=20,40$, and the data shows a clear convergence
towards them. The disjoint case on the right of Fig.~\ref{fig:lnegt} shows qualitatively
similar features, with a delayed increase and decrease of $\lneg(t)$ at times $t=d/2$
and $t=\ell+d/2$, with $d/2$ being the distance from the junction. The precise description
of $\lneg(t)$ between these times is, however, beyond our reach.

\section{Discussion\label{sec:disc}}

We have studied the scaling of entanglement  in a current-carrying NESS of free fermions.
Due to the Gaussianity of the state, both the RMI as well as the R\'enyi generalization
of the entanglement negativity can be obtained via determinants involving the reduced
correlation matrix. Moreover, the analytic continuation in the R\'enyi index can be worked
out explicitly. One finds a logarithmic scaling for both quantities, which is a consequence
of the jump singularity in the NESS occupation function, and the prefactors can be determined
by FH techniques. In particular, we find different prefactors for the negativity as well as for the
RMI with index $\alpha=1/2$.

To derive our analytical result we assumed that the only important details of the otherwise
continuous occupation function are the values at the two jump locations. In our particular case,
the second jump at momentum $q=\pi$ is related to that at $q=0$ by a particle-hole symmetry.
This is equivalent to the change of variables $x_{a,b} \to -x_{a,b}$ and $\tilde x_{a,b} \to - \tilde x_{a,b}$,
which is a trivial symmetry of the prefactors \eqref{sig12} and \eqref{signeg}.
A more general situation would be to consider initial states with different chemical potentials
in the two reservoirs, which would brake the particle-hole symmetry. We conjecture that the
corresponding negativity prefactor would be a sum of the prefactors $\tilde \sigma /2$
evaluated at the corresponding jump parameters, and similarly for the RMI.

Another natural extension of the result would be to consider the NESS of an XY chain,
which is described by block-Toeplitz matrices with similar jump singularities \cite{AP03,Kormos17}.
The result for the RMI can then be obtained \cite{KZ17} by using a generalized FH conjecture
proposed in \cite{AEFQ15}. It would be interesting to extend these techniques to the calculation of the negativity.
Note, however, that for spin models one has the additional difficulty, that the partial transpose
produces a non-Gaussian state \cite{EZ15,CTC15,CTC16}, and the negativities in the spin and fermion representations
are not equivalent. Nevertheless, the quench examples studied so far indicate that these two representations
actually produce very similar results \cite{GE20,MAC22}, at least to leading order. Whether this holds true for
the mixed NESS at hand remains an open question.

In a broader context one might wonder about the origin of the discrepancy between the
$\alpha=1/2$ RMI and the negativity. In the field theory context, this was found to occur only
in irrational CFTs \cite{KFKR20} and was associated to a breakdown of the quasiparticle picture \cite{KFKR21}.
This is clearly not the case here, as the free-fermion chain has perfectly
well defined quasiparticles. However, there is actually no proper CFT description of the
NESS studied here, as the logarithmic scaling of the negativity and RMI is clearly related
to the nonvanishing dispersion at $q \to 0$ and is thus a lattice effect.
On the other hand, discrepancies between $\lneg$ and $\frac{1}{2}\mathcal{I}_{1/2}$
have also been observed in the case of dissipative free-fermion dynamics \cite{AC22,CA22}.
The property this scenario has in common with ours is the global state being mixed.
Whether in such cases the RMI is still a good measure of correlations and provides some
relevant information about the quantum state remains to be further explored.

\begin{acknowledgments}

We thank Z. Zimbor\'as for fruitful discussions and S. Fraenkel and M. Goldstein for useful correspondence.
The author acknowledges funding from the Austrian Science Fund (FWF) through
project No. P35434-N.

\end{acknowledgments}

\newpage

\appendix

\section{Determinant formulas\label{app:det}}

Here we provide determinant formulas that are necessary for the derivation of
Eqs. \eqref{sn} and \eqref{rnegsum} for the R\'enyi entropy and negativity, respectively.
Let us consider a particle-conserving fermionic Gaussian state $\rho$ characterized
by its correlation matrix $C_{mn}=\Tr(\rho \, c_m^\dag c_n)$.
The R\'enyi entropy of a subsystem $A$ is then given by
\eq{
S_n(\rho_A) = \frac{1}{1-n} \ln \det \left[C^n_A + (\identity-C_A)^n\right] ,
\label{snapp}}
where $C_A$ is the reduced correlation matrix and $\identity\equiv\identity_A$.
Using the factorization of the polynomial \eqref{pnz}, the determinant in
\eqref{snapp} can be rewritten as a product
\eq{
\det \left[C^n_A + (\identity-C_A)^n\right] =
\prod_{k=-\frac{n-1}{2}}^{\frac{n-1}{2}}
\det \left[\identity-z_k^{-1}C_A\right] ,
}
where $z_k$ are the roots given in \eqref{zk}. Taking the logarithm and
substituting for $z_k$, one immediately obtains the required formula \eqref{sn}.

The calculation of the negativity is somewhat more complicated. 
In order to evaluate the trace in \eqref{rneg}, it is useful to introduce the
auxiliary density matrix \cite{EEZ18}
\eq{
\rho_\times = \frac{O_+ O_-}{\Tr \, (O_+ O_-)} .
\label{rhox}}
Using multiplication rules for Gaussian states, the corresponding
correlation matrix is given by $C_\times=(\identity+G_\times)/2$,
where $G_\times$ was defined in \eqref{Gx}. The trace of the
$n/2$-th power can then be written as
\eq{
\Tr (\rho_\times)^{\frac{n}{2}} = \det \left[
\left(\frac{\identity+G_\times}{2}\right)^{\frac{n}{2}} +
\left(\frac{\identity-G_\times}{2}\right)^{\frac{n}{2}} \right].
\label{trrhox}}
The normalization factor in \eqref{rhox} is given by
\eq{
\Tr \, (O_+ O_-) = \det \left[\frac{\identity+G_+G_-}{2}\right],
}
where $G_\pm = 2 C_\pm -\identity$.
One then has
\eq{
\lneg_n = \ln \Tr (\rho_\times)^{\frac{n}{2}} +
\frac{n}{2} \ln \Tr \, (O_+ O_-) \, .
\label{lnegn}}
Note that, since \eqref{trrhox} is valid for arbitrary (non-integer) powers,
one can simply substitute $n \to 1$ which yields the formula \eqref{lndet}
in the main text.

The next step is to apply the factorization of the determinant in \eqref{trrhox}.
Using \eqref{tpn} one can rewrite
\eq{
\Tr (\rho_\times)^{\frac{n}{2}}=
\prod_{k=\frac{1}{2}}^{\frac{n-1}{2}}
\det \left[\identity- \tilde z_k^{-1}\frac{\identity-G_\times}{2}\right],
}
where the roots $\tilde z_k$ are defined in \eqref{ztk}. Inserting the expression
\eqref{Gx} of $G_\times$ and taking the logarithm one obtains
\begin{align}
&\ln \Tr (\rho_\times)^{\frac{n}{2}} =
\sum_{k=\frac{1}{2}}^{\frac{n-1}{2}} \left\{
- \ln \det \left[\frac{\identity+G_+G_-}{2}\right]
\right. \nonumber \\
+ & \left. \ln \det \left[\frac{\identity+G_+G_-}{2}-
\frac{\tilde z_k^{-1}}{4}(\identity-G_+)(\identity-G_-)\right] \right\}.
\end{align}
Notice that the first term in the above sum exactly cancels with the
second term in \eqref{lnegn}. One can now apply a similarity transformation
to simplify the remaining determinant. Indeed, one has $G_{\pm}=T_\pm G_A T_\pm$,
where $G_A=2C_A-\identity$ and we introduced the matrices
\eq{
T_\pm =\identity_{A_1} \oplus (\pm i) \identity_{A_2} \, ,
\quad
R =\identity_{A_1} \oplus (- 1) \identity_{A_2} \, .
}
Note that $T_\pm$ and $R$ are diagonal matrices satisfying $T_-=(T_+)^{-1}$
and $(T_\pm)^2=R$, as well as $R^2=\identity$. Using these properties, one arrives at
\eq{
\lneg_n=\sum_{k=\frac{1}{2}}^{\frac{n-1}{2}}
\ln \det \left[\frac{\identity+G_A^2}{2}-
\tilde z_k^{-1}\left(\frac{R-G_A}{2}\right)^2\right].
\label{lnegn2}}

The final step is to factorize the matrix in the above determinant.
Using the expression of the roots \eqref{ztk}, one can check by simple matrix algebra
that the following identity holds \cite{FG22}
\eq{
\frac{\identity+G_A^2}{2}-
\tilde z_k^{-1}\left(\frac{R-G_A}{2}\right)^2 = D^+_k \, \Lambda^{\phantom{+}}_k D^-_k \, ,
}
where $\Lambda_k = \ee^{i(\pi-2\lambda_k) \identity_{A_1}}$ and
\begin{align}
&D^+_k = \identity + C_A (\ee^{i\lambda_k \identity_{A_1}}
\ee^{i (\pi-\lambda_k) \identity_{A_2}}-\identity) \, , \\
&D^-_k = \identity + (\ee^{-i(\pi -\lambda_k) \identity_{A_1}}
\ee^{-i\lambda_k \identity_{A_2}}-\identity) C_A \, .
\end{align}
The determinant in \eqref{lnegn2} thus splits into three parts. 
First we notice that for any even $n$ one has
\eq{
\prod_{k=\frac{1}{2}}^{\frac{n-1}{2}}\det \Lambda_k = 1 \, ,
}
which follows from the half-integer values of $k$. Furthermore,
$\det D^+_k$ is exactly the determinant that appears in \eqref{detzt}.
Finally, let us observe that the phases appearing in $D^-_k$ are both
located on the lower half plane. In fact, one can write
\eq{
\det D^-_k=(\det D^+_{n/2-k})^* \, .
}
Putting everything together, one arrives at the formula \eqref{rnegsum}
of the main text.

\section{Calculation of $\sigma_{1/2}$ and $\tilde\sigma$\label{app:sig}}

In this appendix we calculate the integrals that appear in the prefactors
\eqref{signint} of the R\'enyi mutual information as well as \eqref{signegint}
of the logarithmic negativity. First we focus on $\sigma_{1/2}$ where one has
the following integral
\eq{
\int_a^b  \dd x \, s'_{1/2}(x) \ln \left(\frac{b-x}{x-a}\right).
}
Let us introduce a new variable via
\eq{
\ee^\varepsilon = \sqrt{\frac{1-x}{x}} \, ,
\label{subeps}}
and define the corresponding boundary values accordingly
\eq{
\ee^{\varepsilon_a} = \sqrt{\frac{1-a}{a}} \, , \qquad
\ee^{\varepsilon_b} = \sqrt{\frac{1-b}{b}} \, .
}
Then one has
\eq{
s'_{1/2}(x)=2 \tanh(\varepsilon/2)\cosh(\varepsilon) \, ,
}
while the integration measure transforms as
\eq{
\dd x =  -\frac{1}{2} \frac{\dd \varepsilon}{\cosh^2 (\varepsilon)} \, ,
}
such that the integral can be rewritten as
\eq{
\int \limits_{\varepsilon_b+\delta}^{\varepsilon_a-\delta} \dd \varepsilon
\frac{\tanh(\varepsilon/2)}{\cosh(\varepsilon)}
\ln \left[\frac{\sinh(\varepsilon-\varepsilon_b)\cosh \varepsilon_a}
{\sinh(\varepsilon_a-\varepsilon)\cosh \varepsilon_b}\right] .
}
Note that we introduced an infinitesimal $\delta \to 0$ to regularize the integrand,
which has logarithmic singularities around the boundaries. One can now perform a partial
integration using
\eq{
\frac{\dd}{\dd \varepsilon} \ln \left[
1+\tanh^2(\varepsilon/2) \right]=
\frac{\tanh(\varepsilon/2)}{\cosh(\varepsilon)} \, .
}
The boundary contribution is then
\begin{align}
B_1 &= \ln \left[1+\tanh^2(\varepsilon_a/2)\right] 
\ln \left[\frac{\sinh(\varepsilon_a-\varepsilon_b)\cosh \varepsilon_a}
{ \delta \cosh \varepsilon_b}\right] \nonumber \\
&+\ln \left[1+\tanh^2(\varepsilon_b/2)\right] 
\ln \left[\frac{\sinh(\varepsilon_a-\varepsilon_b) \cosh \varepsilon_b}
{\delta \cosh \varepsilon_a}\right],
\label{B1}
\end{align}
and the remaining integral reads
\eq{
-\int \limits_{\varepsilon_b+\delta}^{\varepsilon_a-\delta} \dd \varepsilon
\ln \left[1+\tanh^2(\varepsilon/2)\right] 
\left[ \coth (\varepsilon-\varepsilon_b) +
\coth (\varepsilon_a-\varepsilon)\right] .
\label{intepsmi}}
We can rewrite
\eq{
\coth (\varepsilon-\varepsilon_b) =
\frac{1}{2}\left[\tanh\Big(\frac{\varepsilon-\varepsilon_b}{2}\Big)+
\coth\Big(\frac{\varepsilon-\varepsilon_b}{2}\Big)\right],
}
and similarly for $\coth (\varepsilon_a-\varepsilon)$.
This suggests another change of variables
\eq{
x = \tanh \left(\frac{\varepsilon}{2}\right) \,,
\label{subx}}
and the boundary values are given by
\begin{align}
&x_a = \tanh\left(\frac{\varepsilon_a}{2}\right)=\frac{\sqrt{1-a}-\sqrt{a}}{\sqrt{1-a}+\sqrt{a}} \,,
\\
&x_b = \tanh\left(\frac{\varepsilon_b}{2}\right)=\frac{\sqrt{1-b}-\sqrt{b}}{\sqrt{1-b}+\sqrt{b}} \,.
\end{align}
Using trigonometric identities, one can rewrite the terms in the integrand of \eqref{intepsmi}
in the new variables as follows
\begin{align}
\tanh\Big(\frac{\varepsilon-\varepsilon_b}{2}\Big) &+
\tanh\Big(\frac{\varepsilon_a-\varepsilon}{2}\Big) =
\frac{(x_a-x_b)(1-x^2)}{(1-x \, x_a)(1-x \, x_b)} \, , \nonumber \\
\coth\Big(\frac{\varepsilon-\varepsilon_b}{2}\Big) &+
\coth\Big(\frac{\varepsilon_a-\varepsilon}{2}\Big) =
\frac{(x_a-x_b)(1-x^2)}{(x - x_b)(x_a - x)} \, .
\end{align}
Noting that
\eq{
\dd x  = \frac{\dd \varepsilon}{2 \cosh^2\frac{\varepsilon}{2}}=\frac{\dd \varepsilon}{2}(1-x^2) \, ,
}
the integral in the new variable thus reads
\begin{align}
-\int \limits_{x_b+\delta'_b}^{x_a-\delta'_a} \dd x 
&\left[\frac{x_a-x_b}{(1-x \, x_a)(1-x \, x_b)} \right . \nonumber \\ & \left.
+ \frac{x_a-x_b}{(x - x_b)(x_a - x)} \right] \ln (1+x^2) \, .
\end{align}
It is important to observe that the infinitesimals $\delta'_a$ and $\delta'_b$ actually depend
on $x_a$ and $x_b$. Indeed, one has
\begin{align}
x_a - \delta'_a &= \tanh\Big(\frac{\varepsilon_a-\delta}{2}\Big)
\approx x_a - \frac{\delta}{2}(1-x_a^2) \, , \nonumber \\
x_b + \delta'_b &= \tanh\Big(\frac{\varepsilon_b+\delta}{2}\Big)
\approx x_b + \frac{\delta}{2}(1-x_b^2) \, .
\label{delta}
\end{align}

Now we perform another partial integration using
\begin{align}
\frac{\dd}{\dd x} \ln \left(\frac{1-x \, x_b}{1-x \, x_a}\right) &=
\frac{x_a-x_b}{(1-x \, x_a)(1-x \, x_b)} \, , \nonumber \\
\frac{\dd}{\dd x} \ln \left(\frac{x- x_b}{x_a-x}\right) &=
\frac{x_a-x_b}{(x - x_b)(x_a - x)} \, .
\end{align}
Using \eqref{delta}, the boundary contribution reads
\begin{align}
B_2 =
&-\ln(1+x^2_a)
\ln \left[\frac{2(1-x_a x_b)(x_a-x_b)}{\delta (1-x^2_a)^2}\right] 
\nonumber \\
&-\ln(1+x^2_b)
\ln \left[\frac{2(1-x_a x_b)(x_a-x_b)}{\delta (1-x^2_b)^2}\right] ,
\label{B2}
\end{align}
and we are left with the integral
\eq{
\int_{x_b}^{x_a} \dd x \frac{2x}{1+ x^2}
\left [\ln \left(\frac{1-x \, x_b}{1-x \, x_a}\right) + 
\ln \left(\frac{x- x_b}{x_a-x}\right) \right].
\label{intxmi}}
Note that the two boundary terms \eqref{B1} and \eqref{B2} can be collected by rewriting
\eq{
\frac{\sinh(\varepsilon_a-\varepsilon_b)\cosh \varepsilon_a}
{\cosh \varepsilon_b}=
\frac{2(1-x_a x_b)(x_a-x_b)}{(1-x^2_a)^2}\frac{1+x^2_a}{1+x^2_b} \, ,
}
and similarly for the other term, such that the overall contribution reads
\eq{
B_1 + B_2= \ln^2 \left(\frac{1+x^2_a}{1+x^2_b}\right) .
\label{B12}}

The integral \eqref{intxmi} can be further simplified by noting that
\eq{
\frac{2x}{1+x^2}=\frac{i}{1+ix}-\frac{i}{1-ix}=
2 \, \rp \left(\frac{i}{1+ix}\right) .
\label{x2re}}
The remaining integrals can be explicitly evaluated in terms of the dilogarithm
function as
\begin{align}
\int_{x_b}^{x_a} \dd x \frac{i}{1+i \, x}
\ln \left(\frac{1-x \, x_b}{1-x \, x_a}\right)= \nonumber \\
-\ln \left(\frac{1+i \, x_a}{1+i \, x_b}\right) \ln \left(\frac{1-i \, x_a}{1-i \, x_b}\right) \nonumber \\
+\Li{\frac{1+ i \, x^{\phantom{-1}}_a}{1+i \, x^{-1}_a}}+\Li{\frac{1+ i \, x^{\phantom{-1}}_b}{1+i \, x^{-1}_b}} \nonumber \\
- \Li{\frac{1+ i \, x^{\phantom{-1}}_b}{1+i \, x^{-1}_a}}-\Li{\frac{1+ i \, x^{\phantom{-1}}_a}{1+i \, x^{-1}_b}},
\label{intx1mi}
\end{align}
while the second integral yields
\eq{
\int_{x_b}^{x_a} \dd x \frac{i}{1+i \, x}
\ln \left(\frac{x- x_b}{x_a-x}\right)=
-\frac{1}{2} \ln^2 \left(\frac{1+i \, x_a}{1+i \, x_b}\right) .
\label{intx2mi}}
Finally, collecting the logarithms in Eqs. \eqref{B12}, \eqref{intx1mi} and \eqref{intx2mi}, with the definition
$z=\frac{1+i \, x_a}{1+i \, x_b}$, one has
\eq{
\ln^2(z \, z^*) - \rp \ln^2(z) - 2\ln(z) \ln(z^*) = \rp \ln^2(z)
}
and one thus arrives at the final result \eqref{sig12} reported in the main text.

The calculation of the negativity prefactor $\tilde \sigma$ follows a very similar line.
The integral that appears in \eqref{signegint} reads
\eq{
\int_{\tilde a}^{\tilde b} \dd x \frac{\tilde p'_1(x)}{\tilde p_1(x)}
\ln \left[\frac{1-a + a \, \sqrt{\frac{x}{1-x}}}{1-b + b \, \sqrt{\frac{x}{1-x}}}\right] .
}
We first observe that
\eq{
\frac{\tilde p'_1(x)}{\tilde p_1(x)} = \frac{1}{2}s'_{1/2}(x) \, .
}
Let us change again variables as in \eqref{subeps}. Note that the integration boundaries
are now different and we define
\eq{
\ee^{\tea} = \sqrt{\frac{1-\tilde a}{\tilde a}} = \frac{1-a}{a} \, , \quad
\ee^{\teb} = \sqrt{\frac{1-\tilde b}{\tilde b}} = \frac{1-b}{b} \, .
}
We then have
\eq{
\frac{1}{2} \int_{\teb}^{\tea} \dd \varepsilon
\frac{\tanh(\frac{\varepsilon}{2})}{\cosh(\varepsilon)}
\ln \left[\frac{\cosh(\frac{\teb}{2})\cosh(\frac{\varepsilon+\tea}{2})}
{\cosh(\frac{\tea}{2}) \cosh (\frac{\varepsilon+\teb}{2})} \right] .
}
We now integrate by parts, with the contribution from the integral boundaries given by
\begin{align}
\tilde B_1&=\frac{1}{2}\ln \left[1+\tanh^2 \Big(\frac{\tea}{2}\Big)\right]
\ln \left[\frac{1+\tanh^2(\frac{\tea}{2})}{1+\tanh(\frac{\tea}{2})\tanh(\frac{\teb}{2})}\right]
\nonumber \\
&+\frac{1}{2}\ln \left[1+\tanh^2\Big(\frac{\teb}{2}\Big)\right]
\ln \left[\frac{1+\tanh^2(\frac{\teb}{2})}{1+\tanh(\frac{\tea}{2})\tanh(\frac{\teb}{2})}\right]
\label{bound1}
\end{align}
whereas the remaining integral to be evaluated reads
\begin{align}
&-\frac{1}{2}\int_{\teb}^{\tea} \dd \varepsilon
\ln \left[ 1+\tanh^2(\varepsilon/2) \right] \nonumber \\ 
&\times \frac{1}{2}
\left[ \tanh \Big(\frac{\varepsilon+\tea}{2}\Big) - 
\tanh \Big( \frac{\varepsilon+\teb}{2} \Big)\right] .
\end{align}
Introducing now the variable change \eqref{subx}, the integral transforms into
\eq{
-\frac{1}{2}\int_{\tilde x_b}^{\tilde x_a} \frac{\dd x}{1-x^2}
\ln \left( 1+x^2 \right)
\left( \frac{x+\tilde x_a}{1+x \, \tilde x_a} - \frac{x+\tilde x_b}{1+x \, \tilde x_b}\right),
\label{intx1}}
where we have defined
\begin{align}
\tilde x_{a}&=\tanh \Big(\frac{\tea}{2}\Big) = 1-2a\, , \nonumber \\
\tilde x_{b}&=\tanh \Big(\frac{\teb}{2}\Big) = 1-2b\, .
\label{xabapp}
\end{align}
The integrand of \eqref{intx1} can be further simplified as
\eq{
-\frac{1}{2}\int_{\tilde x_b}^{\tilde x_a} \dd x
\ln \left( 1+x^2 \right) \frac{\tilde x_a-\tilde x_b}{(1+x \, \tilde x_a)(1+x \, \tilde x_b)} \, .
\label{intx2}}
Now we perform another partial integration using
\eq{
\frac{\dd}{\dd x} \ln \left(\frac{1+x \, \tilde x_a}{1+x \, \tilde x_b}\right)=
\frac{\tilde x_a-\tilde x_b}{(1+x \, \tilde x_a)(1+x \, \tilde x_b)} \, .
}
The boundary contribution then reads
\begin{align}
\tilde B_2=-&\frac{1}{2}\ln (1+\tilde x_a^2) \ln \left(\frac{1+\tilde x_a^2}{1+\tilde x_a \, \tilde x_b}\right) 
\nonumber \\
-&\frac{1}{2}\ln (1+\tilde x_b^2) \ln \left(\frac{1+\tilde x_b^2}{1+\tilde x_a \, \tilde x_b}\right) ,
\label{bound2}
\end{align}
thus using the definition \eqref{xabapp}, it exactly cancels the previous boundary contribution in \eqref{bound1},
$\tilde B_1+\tilde B_2=0$. The remaining integral to be evaluated is then
\eq{
\int_{\tilde x_b}^{\tilde x_a} \dd x
\frac{x}{1+x^2} \ln \left(\frac{1+x \, \tilde x_a}{1+x \, \tilde x_b}\right) .
\label{intx3}}
We use again \eqref{x2re}, such that \eqref{intx3} can be evaluated via the complex valued integral
\begin{align}
&\int_{\tilde x_b}^{\tilde x_a} \dd x \frac{i}{1+ix}
\ln \left(\frac{1+x \, \tilde x_a}{1+x \, \tilde x_b}\right)=
\ln^2 \left(\frac{1+i \, \tilde x_a}{1+i \, \tilde x_b}\right) \nonumber \\
&+ \Li{i \, \tilde x_a\frac{1+ i \, \tilde x_b}{1+i \, \tilde x_a}}+
\Li{i \, \tilde x_b \frac{1+ i \, \tilde x_a}{1+i \, \tilde x_b}} \nonumber \\
&- \Li{i \, \tilde x_a}- \Li{i \, \tilde x_b} 
\end{align}
Finally, using
\eq{
\frac{1-a + a \, i}{1-b + b \, i}=
\frac{1- i \, \tilde x_a}{1- i \, \tilde x_b}
}
in the second piece of \eqref{signegint}, we arrive at the result \eqref{signeg} in the main text.

\bibliography{lognegness_refs}

\end{document}